\documentclass{JHEP3}
\usepackage{amsmath}
\usepackage{epsfig}

\newcommand{\N}{{\cal N}}
\newcommand{\bo}{Bogoliubov }
\newcommand{\HH}{{\cal H}}
\newcommand{\cU}{{\cal U}}
\newcommand{\cS}{{\cal S}}
\newcommand{\cI}{{\cal I}}
\newcommand{\cJ}{{\cal J}}
\newcommand{\s}{{\sigma}}
\newcommand{\w} {\omega}
\newcommand{\ka}{{\kappa}}
\newcommand{\K}{{\rm K}}
\newcommand{\One} {{\bf 1}} 
\newcommand{\tr} {\operatorname{tr}}
\newcommand{\csch} {\operatorname{csch}}
\newcommand{\sech} {\operatorname{sech}}
\newcommand{\threehalves} {{\textstyle\frac{3}{2}}}
\newcommand{\mat}[1] {\begin{pmatrix}#1\end{pmatrix}}
\newcommand{\bra}[1] {\left<#1\right|}
\newcommand{\ket}[1] {\left|#1\right>}
\newcommand{\braket}[2] {\left<#1\vphantom{#2}\right|
                         \left.\!\vphantom{#1}{#2}\right>}

\title{Continuous Half-String Representation of String Field Theory}

\author{Ehud Fuchs, Michael Kroyter, Alon Marcus\\
School of Physics and Astronomy\\
The Raymond and Beverly Sackler Faculty of Exact Sciences\\
Tel Aviv University, Ramat Aviv, 69978, Israel\\
E-mails:
\email{udif@tau.ac.il}, \email{mikroyt@tau.ac.il}, \email{alon@tau.ac.il}
}

\abstract{
We give the explicit form of the half-string representation in the
continuous $\kappa$ basis. We show the comma structure of the three-vertex,
when expanded around an arbitrary projector, and that the zero-mode must be
replaced by the mid-point degree of freedom  in the half-string
representation. The treatment of the ghost sector enables us to calculate
the normalization of the vertices.
The simplicity of this formalism is demonstrated with some applications,
such as gauge transformations and identification of subalgebras.
}

\preprint{TAUP-2715-02\\{\tt hep-th/0307148}}

\begin{document}

\section{Introduction}

Cubic string field theory~\cite{Witten:1986cc} is a candidate for
a non-perturbative, background independent definition of string theory,
which reproduces the perturbative results of string
theory~\cite{Giddings:1986iy}.
Some computational tools were derived in string field theory,
such as mode expansion~\cite{Gross:1987ia,Gross:1987fk,Ohta:1986wn} and CFT
methods~\cite{LeClair:1989sp,LeClair:1989sj},
but the formulation remained complicated.

Following advances in the understanding of non-perturbative string theory,
Sen made his conjectures~\cite{Sen:1999xm} regarding tachyon condensation.
The conjectures were verified using level truncation~\cite{Kostelecky:1990nt}
to a remarkable degree of
accuracy~\cite{Sen:1999nx,Taylor:2000ek,Moeller:2000xv,deMelloKoch:2000ie,
Ellwood:2001py,Taylor:2002fy,Gaiotto:2002wy}.
Yet, there is still no analytic solution describing the
non-perturbative vacuum.

The lack of an analytic solution led Rastelli, Sen and Zwiebach
to propose vacuum string field theory
(VSFT)~\cite{Rastelli:2000hv,Rastelli:2001jb,Rastelli:2001rj,Rastelli:2001uv}.
The solutions of VSFT are projectors \cite{Rastelli:2000hv,Hata:2001sq},
and many such solutions were constructed.
Despite the success of VSFT it has the problem of being
singular~\cite{Gross:2001yk,Gaiotto:2001ji,Moore:2001fg,Erler:2003eq},
and in any case one would like to understand its relation to Witten's SFT.

As a byproduct of the new interest in string field theory, further ways
to simplify the star product were developed, most notably the half-string
formulation~\cite{Bordes:1991xh,Gross:2001rk,Rastelli:2001rj,Furuuchi:2001df},
spectroscopy of the three-vertex~\cite{Rastelli:2001hh,Okuyama:2002yr},
and the Moyal
formulations~\cite{Bars:2001ag,Bars:2002nu,Bars:2003gu,Douglas:2002jm,
Erler:2002nr}.
Different subalgebras of the star product were recognized,
usually belonging to the subalgebra of (shifted) squeezed states.
Among these are the surface states, wedge states 
and $\HH_{\ka^2}$~\cite{Rastelli:2000iu,Fuchs:2002zz}.

The simplification of the star product usually complicates the form
of the BRST operator $Q$.
It may seem that this price is worth paying since $Q$ is limited to
the quadratic term.
However, in order to find the analytic solution, one should deal with
both the star product and $Q$.
In the continuous basis the situation is more complicated,
as it was shown in~\cite{Douglas:2002jm}
that $Q$ cannot be described by a function in this basis.
This problem was partially solved when expressions for $Q$ as a
generalized distribution and as a difference operator were
provided in~\cite{Fuchs:2002wk,Belov:2002te}.

In this paper we present the continuous half-string formalism.
We use the fact that Bogoliubov transformations, which are defined by
squeezed states projectors, transform the full-string into a half-string
basis~\cite{Kostelecky:2000hz}.
Starting from the continuous basis, which is reviewed in
section~\ref{sec:Prel}, we can choose any projector in
$\HH_{\ka^2}$ to get a continuous half-string basis.
The star product and integration then become
\begin{equation}
\begin{gathered}
\label{starForm}
(\Psi\star\Phi)(l^\ka,r^\ka,x_m)=\int Dy^\ka
    \Psi(l^\ka,y^\ka,x_m)\Phi(-y^\ka,r^\ka,x_m)\,,\\
\int\Psi=\int Dy^\ka dx_m\Psi(y^\ka,-y^\ka,x_m)\,,
\end{gathered}
\end{equation}
where $\ka>0$ and the modes $l^\ka,r^\ka$ are linear combinations
of $x^{\pm \ka}$.
This result is independent of the projector we choose to work with,
and it greatly simplifies calculations in the star-algebra.
The mode $x^{\ka=0}$ has no left-right factorization.
It is a single mode in a continuum, and thus it can usually be ignored.
However, states that are not continuous at $\ka=0$ or are
otherwise singular there, can get a nontrivial contribution from this
mode. We shall return to this point at the body of the paper.

The detailed derivation of the continuous half-string basis can be found
in section~\ref{sec:ContHalf}, where we also
calculate explicitly the transformation to the sliver basis and to
the butterfly basis.
The half-string formalism in the sliver basis was studied
in~\cite{Furuuchi:2001df}, while the half-string formalism
of~\cite{Bordes:1991xh}
is actually the half-string in the butterfly basis~\cite{Gaiotto:2002kf}.
We end the section by incorporating the zero mode and the ghost sector
into the formalism, enabling us to calculate the normalization of the
vertices. The infinite factors in the normalization turn out correctly at
the critical dimension $D=26$.

In section~\ref{sec:app} we demonstrate some applications of the formalism.
In~\ref{sec:doubling} we identify a subalgebra which is a doubling of the wedge
state subalgebra.
In~\ref{sec:gauge} we find the explicit form of the gauge transformation
among different butterflies, and in~\ref{sec:tachyon} we show the form of
the Hata Kawano state~\cite{Hata:2001sq} in our representation.
Section~\ref{sec:conclusions} is devoted to conclusions.
The properties of the \bo transformation and its relation to squeezed states
are summarized in the appendix.

\section{Preliminaries}
\label{sec:Prel}

\subsection{The continuous basis}

String field theory has an infinite number of degrees of freedom.
These degrees of freedom describe the string configuration and are usually
represented by the mode expansion of the string configuration and its
conjugate momenta
\begin{equation}
X(\s)=\sqrt{2\alpha'}\sum_{n=0}^\infty x_n u_n(\s)\,,\qquad
P(\s)=\frac{1}{\sqrt{2\alpha'}}\sum_{n=0}^\infty p_n u_n(\s)\,.
\end{equation}
In this notation all the modes are dimensionless,
and the zero modes $x_0,p_0$ are equal to
the physical zero modes $x=\sqrt{2\alpha'}x_0$, $p=p_0/\sqrt{2\alpha'}$
for $2\alpha'=1$.
It is conventional to use $a_0$ instead of what we mark as
$p_0$, yet we use this notation to emphasize that the zero mode is not
an oscillatory mode.

The canonical basis
\begin{equation}
u_0(\s)=1\,,\qquad
u_n(\s)=\sqrt{2}\cos(n\s)\,,
\end{equation}
is complete and orthogonal under the inner product
\begin{equation}
\int_0^\pi \frac{d\s}{\pi} u_n(\s)u_m(\s)=\delta_{nm}\,,\qquad
\frac{1}{\pi}\sum_{n=0}^\infty u_n(\s)u_n(\s')=\delta(\s-\s')\,.
\end{equation}
The creation and annihilation operators
\begin{equation}
\label{aadagger}
a_n=\frac{1}{\sqrt{2}}\left(\frac{p_n}{\sqrt{\w_n}}
    -i\sqrt{\w_n}x_n\right),\qquad
a_n^\dagger=\frac{1}{\sqrt{2}}\left(\frac{p_n}{\sqrt{\w_n}}
    +i\sqrt{\w_n}x_n\right),
\end{equation}
obey the commutation relations $[a_n,a_m^\dagger]=\delta_{nm}$,
and the inverse relations are
\begin{equation}
x_n=\frac{i}{\sqrt{2\w_n}}(a_n-a_n^\dagger)\,,\qquad
p_n=\sqrt{\frac{\w_n}{2}}(a_n+a_n^\dagger)\,.
\end{equation}
By choosing $\w_n=n$ the Hamiltonian gets the canonical form
\begin{equation}
H=\frac{1}{2}\int \frac{d\s}{\pi}
    \left(2\alpha'P^2(\s)+\frac{{X'}^2(\s)}{2\alpha'}\right)
 =\frac{1}{2}\sum_{n=0}^\infty \left(p_n^2+n^2 x_n^2\right)
 =\frac{p_0^2}{2}+\sum_{n=1}^\infty n(a_n^\dagger a_n+\frac{1}{2})\,.
\end{equation}
The normal ordering constant will henceforth be omitted.

The transformation to the continuous $\ka$ basis~\cite{Rastelli:2001hh}
is given by
\begin{equation}
a_\ka=\sum_{n=1}^{\infty} v_n^\ka a_n\,,\qquad
\sum_{n=1}^{\infty} v_n^\ka v_n^{\ka'}=\delta(\ka-\ka')\,,\qquad
\int d\ka v_n^\ka v_m^\ka=\delta_{nm}\,,
\end{equation}
where $v_n^\ka$ are defined by the generating function
\begin{equation}
f_\ka(z)=\sum_{n=1}^\infty\frac{v_n^\ka}{\sqrt{n}}z^n
        =\frac{1}{\sqrt{\N(\ka)}}\frac{1-
e^{-\ka \tan^{-1}z}}{\ka}\,,\qquad
\N(\ka)=\frac{2}{\ka}\sinh\left(\frac{\ka\pi}{2}\right)\,.
\end{equation}
In this basis $\K_1$ is diagonal and the three-vertex is simplified.
It is an orthogonal change of basis for the creation and annihilation
operators that does not alter the vacuum $\ket{\Omega}$
which is annihilated by all $a_n$, and therefore by all $a_\ka$.
The change to the continuous basis does not involve the zero mode,
which remains a discrete degree of freedom.
(A different approach is to diagonalize the three-vertex with the
zero-mode~\cite{Feng:2002rm,Belov:2002fp,Belov:2003df}.)

From the above transformation law we deduce the transformation of the
coordinates
\begin{equation}
\label{xkappa}
\begin{aligned}
x^\ka&=\frac{i}{\sqrt{2\w_\ka}}(a_\ka-a_\ka^\dagger)
     =i\sum_{n=1}^\infty
         \frac{v_n^\ka}{\sqrt{2\w_\ka}}(a_n-a_n^\dagger)
     =\frac{1}{\sqrt{\w_\ka}}\sum_{n=1}^\infty v_n^\ka \sqrt{n}x_n\,,\\
x_n&=\frac{1}{\sqrt{n}}\int d\ka\,v_n^\ka\sqrt{\w_\ka}x^\ka\,.
\end{aligned}
\end{equation}
The eigenstate of the coordinate $x^\ka$ is
\begin{equation}
\label{eigenx}
\ket{x^\ka}=\left(\frac{\w_\ka}{\pi}\right)^{\frac{1}{4}}
  \exp(-\frac{1}{2}\w_\ka (x^\ka)^2-i\sqrt{2\w_\ka}x^\ka a_\ka^\dagger
 +\frac{1}{2}{a_\ka^\dagger}^2)\ket{\Omega}\,.
\end{equation}

For $\w_n$ there is a canonical choice $\w_n=n$,
for which the states generated by acting with creation operators
on the vacuum are eigenstates of the Hamiltonian.
For $\w_\ka$ there is no such canonical choice.
Therefore, we keep this factor explicit in this section, and in the
following sections we set $\w_\ka=1$.
The basis for $X(\s)$ is found from the auxiliary calculation
\begin{multline}
\frac{X(\s)}{\sqrt{2\alpha'}}=\sum_{n=0}^\infty x_n u_n(\s) = x_0+
  \int d\ka\,x^\ka\sqrt{\w_\ka}\sum_{n=1}^\infty\frac{v_n^\ka}{\sqrt{n}}u_n(\s)
\equiv x_0+\int d\ka\,x^\ka u_\ka(\s)\,.
\end{multline}
Notice that we are careful with the positioning of the $\ka$ index in
$x^\ka,u_\ka(\s)$ to distinct the covariant and contravariant vectors.
The generating function can be used to obtain an explicit expression for
the new basis,
\begin{equation}
u_\ka(\s)=\sqrt{\w_\ka}\sum_{n=1}^\infty\frac{v_n^\ka}{\sqrt{n}}u_n(\s)
    =\sqrt{\w_\ka}\sum_{n=1}^\infty\frac{v_n^\ka}{\sqrt{n}}\sqrt{2}\Re e^{in\s}
    =\sqrt{2\w_\ka}\Re f_\ka(e^{i\s})\,,
\end{equation}
which is plotted in figure~\ref{fig:kappa}.
\FIGURE{
\centerline{\begin{picture}(0,0)%
\epsfig{file=Kappa.pstex}%
\end{picture}%
\setlength{\unitlength}{3947sp}%
\begingroup\makeatletter\ifx\SetFigFont\undefined%
\gdef\SetFigFont#1#2#3#4#5{%
  \reset@font\fontsize{#1}{#2pt}%
  \fontfamily{#3}\fontseries{#4}\fontshape{#5}%
  \selectfont}%
\fi\endgroup%
\begin{picture}(6987,1899)(76,-1573)
\put(2701, 89){\makebox(0,0)[lb]{\smash{\SetFigFont{12}{14.4}{\rmdefault}{\mddefault}{\updefault}$\kappa=2$}}}
\put(301, 89){\makebox(0,0)[lb]{\smash{\SetFigFont{12}{14.4}{\rmdefault}{\mddefault}{\updefault}$\kappa=0$}}}
\put(5101, 14){\makebox(0,0)[lb]{\smash{\SetFigFont{12}{14.4}{\rmdefault}{\mddefault}{\updefault}$\kappa=10$}}}
\put( 76,-736){\makebox(0,0)[lb]{\smash{\SetFigFont{12}{14.4}{\rmdefault}{\mddefault}{\updefault}0}}}
\put(2176,-736){\makebox(0,0)[lb]{\smash{\SetFigFont{12}{14.4}{\rmdefault}{\mddefault}{\updefault}$\pi$}}}
\put(2476,-736){\makebox(0,0)[lb]{\smash{\SetFigFont{12}{14.4}{\rmdefault}{\mddefault}{\updefault}0}}}
\put(4576,-736){\makebox(0,0)[lb]{\smash{\SetFigFont{12}{14.4}{\rmdefault}{\mddefault}{\updefault}$\pi$}}}
\put(4876,-736){\makebox(0,0)[lb]{\smash{\SetFigFont{12}{14.4}{\rmdefault}{\mddefault}{\updefault}0}}}
\put(6976,-736){\makebox(0,0)[lb]{\smash{\SetFigFont{12}{14.4}{\rmdefault}{\mddefault}{\updefault}$\pi$}}}
\end{picture}
}
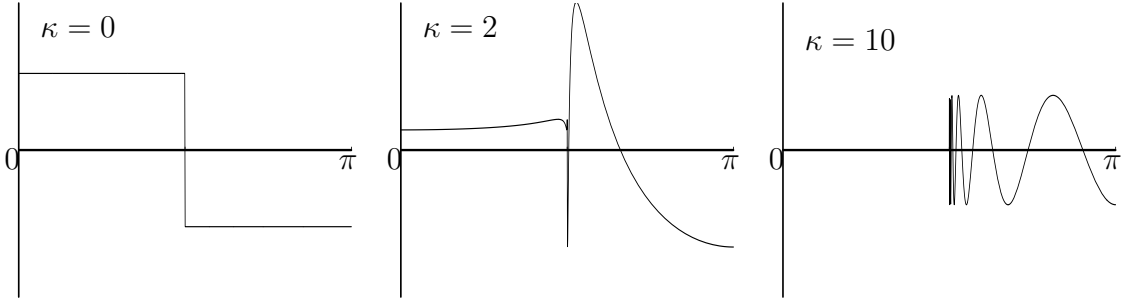
\caption{ The shape of the string basis functions.
For $\kappa=0$ we have a step function around the string mid-point
$\s=\pi/2$. From the picture it is obvious why this state is twist odd.
For other values of $\ka$ these functions are infinitely oscillatory near
the mid-point as described in~\cite{Erler:2003eq}.}
\label{fig:kappa}
}
These basis functions are not orthogonal and
in~\cite{Fuchs:2002wk,Belov:2002te} their inner product was found to give
the metric
\begin{multline}
g_{\ka\ka'}=\int \frac{d\s}{\pi} u_\ka(\s) u_{\ka'}(\s)=
\sqrt{\w_\ka \w_{\ka'}}\sum_{n=1}^\infty \frac{1}{n}v_n^\ka v_n^{\ka'}=\\
 \sqrt{\w_\ka \w_\ka'}\frac{2\sinh(\frac{\ka \pi}{4})\sinh(\frac{\ka' \pi}{4})}
     {\sqrt{\N(\ka)\N(\ka')}\ka \ka' \cosh(\frac{(\ka-\ka') \pi}{4})}
  =\sqrt{\w_\ka \w_\ka'}
     \frac{\sqrt{\tanh(\frac{\ka \pi}{4})\tanh(\frac{\ka' \pi}{4})}}
     {2\sqrt{\ka \ka'} \cosh(\frac{(\ka-\ka') \pi}{4})}\,.
\end{multline}
Repeating the calculation for the conjugate momenta gives
\begin{align}
p_\ka&=\sqrt{\w_\ka}\sum_{n=1}^\infty\frac{v_n^\ka}{\sqrt{n}}p_n\,,\\
u^\ka(\s)&=\frac{1}{\sqrt{\w_\ka}}\sum_{n=1}^\infty\sqrt{n}v_n^\ka u_n(\s)
         =\frac{1}{\sqrt{\w_\ka}}\sqrt{2}
          \Re\left.z\frac{d}{dz}f_\ka(z)\right|_{z=e^{i\s}}\,.
\end{align}
The (very singular) inverse metric was calculated
in~\cite{Fuchs:2002wk,Belov:2002te}
\begin{equation}
\begin{aligned}
g^{\ka\ka'}=&\int \frac{d\s}{\pi} u^\ka(\s) u^{\ka'}(\s)=
  \frac{1}{\sqrt{\w_\ka \w_{\ka'}}}
    \sum_{n=1}^\infty n v_n^\ka v_n^{\ka'}\\
=&\frac{\sqrt{\ka\ka'}}{4\sqrt{\w_{\ka}\w_{\ka'}}}
  (\delta(\ka-\ka'+2i)+\delta(\ka-\ka'-2i))\,.
\end{aligned}
\end{equation}
This expression is very useful, since it gives the expression for
the Hamiltonian in the $\ka$ basis
\begin{equation}
\begin{aligned}
H&=\sum_{n=1}^\infty n a_n^\dagger a_n
=\int d\ka d\ka'\sum_{n=1}^\infty n v_n^\ka v_n^{\ka'} a_\ka^\dagger a_{\ka'}\\
 &=\int d\ka d\ka' \sqrt{\w_\ka \w_\ka'}g^{\ka\ka'}
   a_\ka^\dagger a_{\ka'}\,,
\end{aligned}
\end{equation}
which is $L_0$ and is also
the matter part of $Q_B$ in the Siegel gauge.

Finally, we write the three-vertex in the continuous basis
\begin{equation}
\label{V3full}
\ket{V_3(P_0)}=
\delta(p_0^1+p_0^2+p_0^3)
    \exp\left(-\frac{1}{2}A^\dagger V_3 A^\dagger
    +P_0 V_0 A^\dagger-\frac{1}{2}V_{00}P_0^2\right)\ket{\Omega}_{123}\,,
\end{equation}
where $A^\dagger,V_3,V_0,P_0$ are tensors on the three Fock spaces
\begin{align}
A^\dagger=\mat{a_1^\dagger \\ a_2^\dagger \\ a_3^\dagger},\qquad
V_3=\mat{V_3^{11} & V_3^{12} & V_3^{21}\\V_3^{21}&V_3^{11}&V_3^{12}\\
    V_3^{12}&V_3^{21}&V_3^{11}},\qquad
V_0=\mat{V_0^{11} & V_0^{12} & V_0^{21}\\V_0^{21}&V_0^{11}&V_0^{12}\\
         V_0^{12}&V_0^{21}&V_0^{11}},\qquad
P_0=\mat{p_0^1\\p_0^2\\p_0^3},
\end{align}
and an integration over $\ka$ is assumed. It is convenient to pair up
creation operators which have the same $(\K_1)^2$ eigenvalue
\begin{equation}
\label{adaggervector}
a^\dagger=\mat{a_{-\ka}^\dagger\\a_\ka^\dagger}.
\end{equation}
In this notation we choose $\ka>0$, while $\ka=0$ has to be treated separately.
The three-vertex coefficients in this notation are
\begin{equation}
\label{V3}
\begin{split}
V_3^{11}&=\frac{1}{1+2\cosh(\frac{\ka\pi}{2})}
\mat{0 & 1 \\
     1 & 0},\\
V_3^{12}&=\frac{-1}{1+2\cosh(\frac{\ka\pi}{2})}
\mat{0 & 1+\exp(\frac{\ka\pi}{2}) \\
     1+\exp(-\frac{\ka\pi}{2}) & 0},\\
V_3^{21}&=(V_3^{12})^T\,,\\
V_0^{11}&=\sqrt{2\alpha'}\frac{1}{3}
    \frac{\sqrt{\N(\ka)}\tanh(\frac{\ka\pi}{4})}{1+2\cosh(\frac{\ka\pi}{2})}
    \mat{-1\\1},\\
V_0^{12}&=\sqrt{2\alpha'}\frac{1}{2}
    \frac{\sqrt{\N(\ka)}}{1+2\cosh(\frac{\ka\pi}{2})}
    \mat{1\\1}-\frac{1}{2}V_0^{11}\,,\\
V_0^{21}&=\sqrt{2\alpha'}\frac{1}{2}
    \frac{\sqrt{\N(\ka)}}{1+2\cosh(\frac{\ka\pi}{2})}
    \mat{-1\\-1}-\frac{1}{2}V_0^{11}\,,\\
V_{00}&=(2\alpha')\frac{1}{2}\log(\frac{27}{16})\,.
\end{split}
\end{equation}
From the form of the three-vertex, it can be seen that squeezed states,
whose defining matrix is block diagonal with two-by-two blocks, which mix
$a_\ka$ with $a_{-\ka}$, form a subalgebra.
This is the $\HH_{\ka^2}$ subalgebra of~\cite{Fuchs:2002zz}.

For $\ka=0$ the contribution to the three-vertex is
\begin{equation}
\begin{gathered}
\ket{V_3^{\ka=0}}=
    \exp\left(-\frac{1}{2}A_{\ka=0}^\dagger V_3^{\ka=0}A_{\ka=0}^\dagger
              +P_0 V_0^{\ka=0}A_{\ka=0}^\dagger\right)
    \ket{\Omega},\\
V_3^{\ka=0}=\frac{1}{3}\mat{1&-2&-2\\-2&1&-2\\-2&-2&1},\qquad
V_0^{\ka=0}=\sqrt{2\alpha'}\frac{\sqrt{\pi}}{6}\mat{0&1&-1\\-1&0&1\\1&-1&0}.
\end{gathered}
\end{equation}
In the functional basis, using~(\ref{eigenx}), the zero momentum sector
gives a factor of\footnote{
We have to be careful in our normalization since the three-vertex in the
functional basis is not dimensionless. In the continuous half-string basis
it will be evident that the normalization between the three-vertex state
and functional can be naturally set to 1 by
taking $\w_\ka^l=\w_\ka^r$, and in any case the normalization is
dimensionless since the dimensions of the three-vertex are contained in
the $\delta$-functions.
This pairing of modes fails for $\ka=0$, but
of course, dependence on $\w_0$ should disappear in all final results.
}
\begin{equation}
\label{ka0Vertex}
V_3^{\ka=0}(x^{\ka=0}_{1,2,3})=
\bra{x^{\ka=0}_1}\bra{x^{\ka=0}_2}\braket{x^{\ka=0}_3}{V_3^{\ka=0}}=
\frac{\sqrt{3}}{2}\left(\frac{\w_0}{\pi}\right)^{1/4}
\delta(x_1^{\ka=0}+x_2^{\ka=0}+x_3^{\ka=0})\,.
\end{equation}
This is expected, since
$\ka=0$ describes a step in the middle of the string, as can be seen
in figure~\ref{fig:kappa}.
The $\ka=0$ eigenvalue may seem to be negligible, since it is a
single point in a continuous set of states.
However, there are many calculations in the literature in which
the opposite is true.
This can happen when the state has some kind of a singularity
near $\ka=0$.
In~\cite{Hata:2001wa} this effect was called twist anomaly.
A careful treatment of twist anomaly gave the correct ratio of
D-brane tensions~\cite{Okuyama:2002tw} and the correct tachyon mass
around VSFT's D-brane~\cite{Hata:2002it}.
Such a treatment involves a regularization of the states in hand,
or a regularization of the vertex.
We will demonstrate how the tachyon state of~\cite{Hata:2002it} can
be regularized in the continuous half-string basis, to reproduce the
tachyon's mass. This and other calculations do not require a special
treatment of the $\ka=0$ mode. Therefore, 
we shall not need eq.~\ref{ka0Vertex}.

The importance of $\ka=0$ can also be seen in the discrete Moyal
formalism~\cite{Bars:2001ag,Bars:2002nu}.
Without it, $L_0$ has the wrong spectrum.
Instead of having eigenstates for all integer eigenvalues,
one gets eigenstates with even integer eigenvalues only,
which are doubly degenerate.
This is the spectrum of two strings with half the length of the original
string.
Only when regularizing $L_0$
to include $\ka=0$, the correct spectrum is obtained~\cite{Bars:2003cr}.
Also, when disregarding $\ka=0$, it turns out that the butterfly solves
the equation of motion of string field theory in the Siegel gauge.
This is due to the fact that the butterfly state is the vacuum state of
the two halves of the string and without $\ka=0$, string field theory
essentially becomes a free theory for the two halves of the string.
Ignoring $\ka=0$ in the Lagrangian allows for an arbitrary
value for $x_{\ka=0}$, giving string configurations in which
the string splits in the middle.

\subsection{Half-string representation}

The half-string representation was suggested already in~\cite{Witten:1986cc}
as a convenient representation for string field theory.
We split the string into left and right degrees of freedom
\begin{equation}
X(\s)=l(\s)+\bar{r}(\s)+x_m\,,
\end{equation}
where $\bar{r}(\s)\equiv r(\pi-\s)$, so that both $l(\s)$ and $r(\s)$ are
defined between $0$ and $\pi/2$ and vanish between $\pi/2$ and $\pi$.
The mid-point degree of freedom $x_m\equiv X(\pi/2)$ appears, because we
chose Dirichlet boundary condition for $l(\s),r(\s)$ at the mid-point.

The integration and star product can now be written as
\begin{align}
\int \Psi &= \int Dy(\s)dx_m\Psi(y,y,x_m)\,,\\
(\Psi_1\star\Psi_2)(l,r,x_m)&=\int Dy(\s)\Psi_1(l,y,x_m)\Psi_2(y,r,x_m)\,.
\end{align}
In this representation the string field can be treated as a matrix
over the space of half-strings, up to mid-point issues.

In~\cite{Bordes:1991xh} string field theory was formulated by expanding
the half-string in modes
\begin{align}
l(\s)=\sqrt{2}\sum_{n=1}^\infty l_n \cos(2n-1)\s\,,\\
r(\s)=\sqrt{2}\sum_{n=1}^\infty r_n \cos(2n-1)\s\,.
\end{align}
We relate creation and annihilation operators to these modes as in the case
of the full-string modes~(\ref{aadagger}).
Note that the vacuum state $\ket{\Omega_h}$,
which is annihilated by all the half-string
annihilation operators $a^{l,r}_n$,
is not the same as the vacuum state $\ket{\Omega}$
annihilated by the full-string oscillators $a_n$.

The three-vertex has a simple functional form in this basis,
\begin{equation}
\label{v3hFunc}
V_3^h(l_n^{1,2,3},r_n^{1,2,3},x_m^{1,2,3})=
    \delta(x_m^1-x_m^2)\delta(x_m^2-x_m^3)
  \prod_n
    \delta(r_n^1-l_n^2)\delta(r_n^2-l_n^3)\delta(r_n^3-l_n^1)\,,
\end{equation}
and we can use the state formalism~(\ref{eigenx}) to calculate the
three-vertex state
\begin{equation}
\begin{aligned}
\label{v3hvertex}
\ket{V_3^h(p_m^{1,2,3})}&=\int dx_m^{1,2,3}
e^{-i\sum_{s=1}^3 p_m^sx_m^s}
   \prod_n\left(dl_n^{1,2,3}dr_n^{1,2,3}\right)
   V_3^h(r_n^{1,2,3},l_n^{1,2,3},x_m^{1,2,3})
\ket{l_n^{1,2,3},r_n^{1,2,3}}\\
 &=
\delta(p_m^1+p_m^2+p_m^3)
\exp\left(-\frac{1}{2}\sum_{s,t=1}^3
\sum_{n,m=1}^\infty
(a_h^\dagger)_n^s (V_3^h)_{nm}^{st} (a_h^\dagger)_m^t\right)
\ket{\Omega_h}_{123}\,,
\end{aligned}
\end{equation}
where ${a_h}_n=({a_l}_n,{a_r}_n)$, and $V_3^h$ is composed of the matrices
\begin{equation}
\label{v3hnm}
\begin{aligned}
(V_3^h)_{nm}^{11}=(V_3^h)_{nm}^{22}=(V_3^h)_{nm}^{33}&=\mat{0&0\\0&0}\,,\\
(V_3^h)_{nm}^{12}=(V_3^h)_{nm}^{23}=(V_3^h)_{nm}^{31}&=
    \mat{0&\delta_{nm}\\0&0}\,,\\
(V_3^h)_{nm}^{21}=(V_3^h)_{nm}^{32}=(V_3^h)_{nm}^{13}&=
    \mat{0&0\\\delta_{nm}&0}\,.
\end{aligned}
\end{equation}
Notice that the vertex in~(\ref{v3hFunc}) multiplies states of the
form $\Psi[l,r]=\braket{l,r}{\Psi}$.
Therefore, to calculate $\ket{V_3^h}$, we used the BPZ conjugate of the vertex
\begin{equation}
bpz\left(V_3^h(l_n^{1,2,3},r_n^{1,2,3},x_m^{1,2,3})\right)
   =V_3^h(r_n^{1,2,3},l_n^{1,2,3},x_m^{1,2,3})\,.
\end{equation}

\section{Continuous half-string representation}
\label{sec:ContHalf}

In this section we derive the form of the vertex in the continuous half-string
representation. We start with the $p=0$ matter sector. We then add the
zero mode, which is replaced by the mid-point degree of freedom. Finally,
we discuss the ghost sector, which enables us to calculate the normalization
of the vertices.

\subsection{Zero momentum}
\label{sec:p=0}

In~\cite{Bordes:1991xh} the half-string representation was described
using discrete half-string oscillator.
We suggest using a continuous set of oscillators.
We use the fact that the form of the three-vertex for the half-string
should always be as in eq.~(\ref{v3hFunc}).
Eq.~(\ref{v3hvertex}),(\ref{v3hnm}) then imply
that the half-string vacuum state always obeys
\begin{equation}
\ket{\Omega_h\star\Omega_h}_3=
\vphantom{\bra{\Omega_h}}_1
\!\!\bra{\Omega_h}\vphantom{\bra{Omega_h}}_2
\!\!\bra{\Omega_h}|V_3^h\rangle_{123}
    =\exp(-\frac{1}{2}a_3^\dagger V^{33}a_3^\dagger)\ket{\Omega_h}_3
    =\ket{\Omega_h}_3\,,
\end{equation}
meaning that this vacuum state is a projector.
Here, we assumed that the vacuum state is BPZ real,
meaning $(\ket{\Omega_h})^\dagger=\bra{\Omega_h}$.
We can now find half-string representations using the fact that
every squeezed state projector describes a Bogoliubov
transformation into a half-string basis
(see appendix~\ref{sec:Bogo} for details on the Bogoliubov transformation).
For a given squeezed state matrix $S$, we can find $W$ that satisfies
$W^\dagger W=(1-S S^*)^{-1}$. 
If $S$ describes a projector, then the set of operators
\begin{equation}
\label{aHalf}
a^h_n=W_{nm}a_m+U_{nm}a_m^\dagger\,,\qquad
{a^h_n}^\dagger=W^{*}_{nm}a_m^\dagger+U^{*}_{nm}a_m\,,
\end{equation}
where $U=WS$,
describe oscillators of the half-string for which the squeezed state
is the new vacuum. Remember that $S$ defines $W$ only up to a
unitary transformation. Therefore, this half-string basis could
contain a mixture of left and right modes, meaning that we still
do not have the left-right factorization.

The representation of any squeezed state in the half-string basis can be
obtained using the expressions in~\ref{sec:TofStates}, where we think of
the squeezed state as defining another \bo transformation.
The three-vertex is a (singular) squeezed state in a tripled tensor space
of oscillators.
Therefore, the three-vertex in the new half-string basis is
\begin{equation}
V_3^h=\mat{
V^h_{11} &V^h_{12} &V^h_{21} \\
V^h_{21} &V^h_{11} &V^h_{12} \\
V^h_{12} &V^h_{21} &V^h_{11} }=
  (W_3^\dagger-V_3 U_3^\dagger)^{-1}(V_3W_3^T-U_3^T)\,,
\end{equation}
where $W_3=\One_3 \otimes W$ and $U_3=\One_3 \otimes U$.

For states in $\HH_{\ka^2}$~\cite{Fuchs:2002zz},
we can calculate $V_3^h$ explicitly.
A general state in $\HH_{\ka^2}$ is a squeezed state of the form
$\exp(-\frac{1}{2}a^\dagger S a^\dagger)\ket{\Omega}$,
where $a^\dagger$ is defined in~(\ref{adaggervector}).
If the state is BPZ real, then $S$ can be represented by the matrix
\begin{equation}
\label{SForm}
S=\mat{s_1+is_3&s_2\\s_2&s_1-is_3}=s_1\One+s_2\s_1+is_3\s_3\,,
\end{equation}
where $\s_i$ are the Pauli matrices and $s_i$ are real.

$W$ is defined through~(\ref{WofS}) up to a unitary transformation.
We choose to work with a Hermitian $W$.
There are still four such choices related to the choice of the sign
of the two eigenvalues. We choose $W$ with positive eigenvalues, as
in~(\ref{WHermitian}).
To calculate $W$ we write
\begin{equation}
1-SS^*=A(\One\cosh\alpha+\hat\s\sinh\alpha)=A\exp(\alpha\hat\s)\,,
\end{equation}
where
\begin{equation}
\begin{gathered}
A^2=(1-s_1^2-s_2^2-s_3^2)^2-4s_2^2(s_1^2+s_3^2)\,,\\
\hat\s=\frac{s_1\s_1-s_3\s_2}{\sqrt{s_1^2+s_3^2}}\qquad({\hat\s}^2=1)\,,\\
\alpha=\tanh^{-1}\left(\frac{-2s_2\sqrt{s_1^2+s_3^2}}{1-s_1^2-s_2^2-s_3^2}\right)\,,
\end{gathered}
\end{equation}
and we get
\begin{equation}
W={(1-SS^*)^{-1/2}}=A^{-1/2}\left(\One\cosh(\frac{\alpha}{2})
    -{\hat\s}\sinh(\frac{\alpha}{2})\right)\,.
\end{equation}
Taking a state $S$, which is a projector, we get for the transformed
three-vertex
\begin{equation}
\label{V3Comma}
V^h_{11}=\mat{0 & 0\\0 & 0}\,,\quad
V^h_{12}=-\mat{0 & 1\\0 & 0}\,,\quad
V^h_{21}=-\mat{0 & 0\\1 & 0}\,.
\end{equation}
From the form of $V_3^h$, we deduce that our choice of Hermitian
$W$ gives the correct half-string factorization,
\begin{equation}
a^h_\ka=\mat{a^l_\ka\\a^r_\ka}=
W\mat{a_{-\ka}\\a_\ka}+U\mat{a^\dagger_{-\ka}\\a^\dagger_\ka}.
\end{equation}
Moreover, the
condition on $S$ that the three-vertex will be of the form~(\ref{V3Comma}),
is exactly the condition found in~\cite{Fuchs:2002zz} for $S$ to be a
projector. We do not get the identity solution $S=C$ this way, because
it is of infinite rank, while only rank one projectors have left--right
factorization.

The sign of the $V_3^h$ matrices is opposite to that of the usual
half-string vertex~(\ref{v3hnm}).
This sign can be reversed by taking $W$ with one positive and one negative
eigenvalue, but we chose to work with the positively defined $W$.
In this convention we get that the twist matrix transforms to itself
for any real $S$,
\begin{equation}
C_h=(W-CU)^{-1}(CW-U)=C=-\mat{0&1\\1&0}.
\end{equation}

We can get the three-vertex in the functional form using the expression
for eigenstates of the coordinate $x^\ka$~(\ref{eigenx}),
where our coordinates are $l^\ka,r^\ka$. The three-vertex becomes
\begin{equation}
\label{v3hkafunc}
V_3^h(l^\ka_{1,2,3},r^\ka_{1,2,3})=
   \delta(r^\ka_1+l^\ka_2)\delta(r^\ka_2+l^\ka_3)\delta(r^\ka_3+l^\ka_1)\,,
\end{equation}
where a product over all continuous values of $\ka>0$ is assumed.
The signs in this three-vertex functional are not the standard
signs used as a result of the different sign in the three-vertex state.
We will see that, nonetheless, they give the correct gluing of the strings.

Our results are true for all projectors, but for the sake of simplicity,
we focus on the sliver and the butterfly
\begin{equation}
\begin{aligned}
S_s&=e^{-\frac{\ka\pi}{2}}\mat{0&1\\1&0},&
W_s&=\frac{1}{\sqrt{1-e^{-\ka\pi}}}\mat{1&0\\0&1},&
U_s&=\frac{e^{-\frac{\ka\pi}{2}}}{\sqrt{1-e^{-\ka\pi}}}\mat{0&1\\1&0},\\
S_b&=\frac{1}{2\cosh(\frac{\ka\pi}{2})}\mat{1&1\\1&1},\quad\!&
W_b&=\frac{1}{2\sinh(\frac{\ka\pi}{2})}\mat{
    e^{\frac{\ka\pi}{2}} & e^{-\frac{\ka\pi}{2}}\\
    e^{-\frac{\ka\pi}{2}} & e^{\frac{\ka\pi}{2}}},\quad\!&
U_b&=\frac{1}{2\sinh(\frac{\ka\pi}{2})}\mat{1&1\\1&1}.
\end{aligned}
\end{equation}

\subsection{The shape of the half-string modes}

We can use the relation between the Bogoliubov transformation and coordinate
transformation~(\ref{coorTrans}) to get the shape of the half-string modes
\begin{align}
\label{CoTrans}
\tilde u_\ka^h(\s)&=\mat{\tilde u_\ka^l(\s) \\ \tilde u_\ka^r(\s)}=
    (W+U)\mat{u_{-\ka}(\s) \\ u_\ka(\s)}\,,\\
\label{ContraTrans}
\tilde u^\ka_h(\s)&=\mat{\tilde u^\ka_l(\s) \\ \tilde u^\ka_r(\s)}=
    (W-U)\mat{u^{-\ka}(\s) \\ u^\ka(\s)}\,.
\end{align}
The tilde over $u_\ka^h(\s)$ marks that this result does not include the
zero mode, which will be treated in the next section.
A direct analysis shows that $\tilde u_\ka^l(\s)$, which is related
to the left half-string shape in position space, is constant for $\s>\pi/2$,
as can be seen in figure~\ref{fig:lr}.
The fact that this constant is not zero is consistent, since we should have
added a zero-mode that cancels this constant, and indeed in the next section
this constant will have an important role.
The functions $\tilde u^\ka_l(\s)$ are related to the left half-string
momentum modes, and vanish for $\s>\pi/2$.
Here, the constant vanishes since we are working in the zero momentum sector.
\FIGURE{
\centerline{\begin{picture}(0,0)%
\epsfig{file=lr.pstex}%
\end{picture}%
\setlength{\unitlength}{3947sp}%
\begingroup\makeatletter\ifx\SetFigFont\undefined%
\gdef\SetFigFont#1#2#3#4#5{%
  \reset@font\fontsize{#1}{#2pt}%
  \fontfamily{#3}\fontseries{#4}\fontshape{#5}%
  \selectfont}%
\fi\endgroup%
\begin{picture}(6525,1797)(451,-1546)
\put(451,-736){\makebox(0,0)[lb]{\smash{\SetFigFont{12}{14.4}{\rmdefault}{\mddefault}{\updefault}0}}}
\put(3526,-661){\makebox(0,0)[lb]{\smash{\SetFigFont{12}{14.4}{\rmdefault}{\mddefault}{\updefault}$\pi$}}}
\put(3901,-661){\makebox(0,0)[lb]{\smash{\SetFigFont{12}{14.4}{\rmdefault}{\mddefault}{\updefault}0}}}
\put(6976,-661){\makebox(0,0)[lb]{\smash{\SetFigFont{12}{14.4}{\rmdefault}{\mddefault}{\updefault}$\pi$}}}
\end{picture}
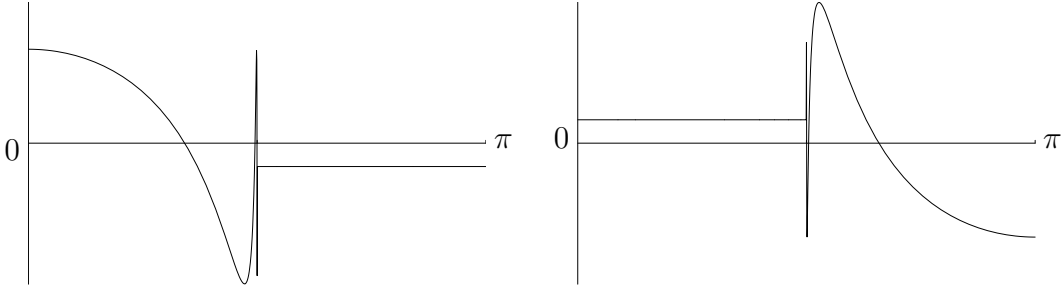
} \caption{ The shape of the string
base functions. $\tilde u_\ka^l(\s),\tilde u_\ka^r(\s)$ are
plotted for $\ka=2$. They are related to each other by a twist around $\pi/2$
and a change of sign. The change of sign explains the signs in the
$\delta$-functions of (\ref{v3hkafunc}).}
\label{fig:lr}
}

\subsection{The zero-mode}

In all the calculations so far, the zero-mode was ignored by setting
$p=0$. In this section we reinstate it in the half-string formalism.
The method of calculating the three-vertex is the same. We treat it
as a Bogoliubov transformation. The difference is that now, in addition
to the term quadratic in creation operators, we have a linear term 
which depends on the zero-mode momenta.
We treat the momenta as parameters of a shifted Bogoliubov transformation.
The calculation results in a diverging coefficient
for the $p_0^2$ term.

We can avoid this divergence by a change of variables from the zero-mode
to the mid-point. The mid-point degree of freedom is defined as
\begin{align}
\label{MidPoint}
x_m&=\frac{X(\pi/2)}{\sqrt{2\alpha'}}
    =x_0+\sum_{n=1}^\infty J_n x_n=x_0+\int d\ka J_\ka x^\ka\\
J_{2n}&\equiv \sqrt{2}(-1)^n\qquad J_{2n+1}\equiv 0\\
J_\ka&=\sum_{n=1}^\infty\frac{v_n^\ka}{\sqrt{n}}J_n=\sqrt{2}\Re f_\ka(i)=
{\cal P}\frac{\sqrt{2}}{\ka\sqrt{\N(\ka)}}\,,
\end{align}
where $J_\ka$ should be regarded as a distribution.
We transform $J_\ka$ to the half-string basis using~(\ref{CoTrans})
\begin{equation}
J_\ka^h=\mat{J_\ka^l\\J_\ka^r}=(W+U)\mat{J_{-\ka}\\J_\ka}.
\end{equation}
In the case of the butterfly basis we get that $J_\ka^h=J_\ka$.
$J_\ka^h$ is exactly the constant value of $u_\ka^h(\s)$
\begin{equation}
\tilde u_\ka^l(\pi/2<\s<\pi)=J_\ka^l\qquad
\tilde u_\ka^r(0<\s<\pi/2)=J_\ka^r\,.
\end{equation}

The expansion of the string configuration with the zero-mode and then with the
mid-point is
\begin{equation}
\label{FieldWithZero}
\frac{X(\s)}{\sqrt{2\alpha'}}
   =x_0+\int d\ka\left(l^\ka \tilde u^l_\ka(\s)+r^\ka \tilde u^r_\ka(\s)\right)
   =x_m+\int d\ka \left(l^\ka u^l_\ka(\s)+r^\ka u^r_\ka(\s)\right),
\end{equation}
where $u_\ka^h(\s)=\tilde u_\ka^h(\s)-J_\ka^h$.
We find the related Bogoliubov transformation from the transformation
of the coordinates
\begin{equation}
\mat{x_m\\h^\ka}=\mat{1&{J_\ka}^T\\0&W-U}\mat{x_0\\x^\ka}\equiv M\mat{x_0\\x^\ka}\,.
\end{equation}
We use a vector notation where the first entry represents a single degree
of freedom (the mid-point or the zero mode) and the second entry represents
a continuous infinite set of degrees of freedom (left and right of the
half-string or $\pm\ka$ of the full-string).
$M$ is an infinite matrix, but it can be easily inverted 
since we have $(W-U)^{-1}=W+U$ and $J^h_\ka=(W+U)J_\ka$ to give
\begin{equation}
M^{-1}=\mat{1&-{J^h_\ka}^T\\0&W+U}\,.
\end{equation}
The transpose of this matrix transforms the momenta and specifically,
for the mid-point momenta, we get $p_m=p_0$.
The transformation of the creation operators is then
\begin{equation}
\label{creationTrans}
a_h^\dagger=W a_\ka^\dagger+U a_\ka-\frac{1}{\sqrt{2}}J_\ka^h p_0\,.
\end{equation}

We can now use the results of~\ref{sec:TofStates} to transform the
three-vertex~(\ref{V3full}). Setting $V=V_3$, $\mu=P_0V_0$ we find the
three-vertex of the continuous half-string including the mid-point,
\begin{multline}
\label{v3hderive}
\ket{V_3^h(P_m)}=
\delta(p_m^1+p_m^2+p_m^3)\cdot\\
\exp\left(-\frac{1}{2}V_{00} P_m^2
+\frac{1}{2}\mu^T(1+V_3)^{-1}\mu-\frac{1}{2}\hat\mu^T(1+V_3^h)^{-1}\hat\mu
+\hat\mu A_h^\dagger-\frac{1}{2}A_h^\dagger V_3^h A_h^\dagger
\right)\ket{\Omega_h}_{123},
\end{multline}
where $A_h=(a_h^1,a_h^2,a_h^3), P_m=(p_m^1,p_m^2,p_m^3)$.
A direct calculation shows that after applying the $\delta$-function
over the momenta, we get $\hat\mu=0$ as expected.
In the derivation of (\ref{v3hderive}) we have replaced
\begin{equation}
(1-V_3)(1-V_3^2)^{-1}=(1+V_3)^{-1}\,.
\end{equation}
This is supposedly forbidden
since $V_3$ has the eigenvalue $1$ in its spectrum. The result may also seem
meaningless because the other eigenvalue of $V_3$ is $-1$.
However, $\mu$ turns out to be an eigenvector of $V_3$ with the eigenvalue $1$.
(Notice that $\mu$ is $P_0$ dependent, but it is an eigenvector of $V_3$
for all values of $P_0$.)
We are left with the term $\frac{1}{4}\mu^T\mu$, which can be easily
calculated in the $\ka$ basis, and after applying the $\delta$ function
over the momenta,
this term exactly cancels $V_{00}$ so that we are left
with the expected three-vertex
\begin{equation}
\label{V3matter}
\ket{V_3^h(P_m)}=
\delta(p_m^1+p_m^2+p_m^3)
\exp\left(-\frac{1}{2}A_h^\dagger V_3^h A_h^\dagger\right)
    \ket{\Omega_h}_{123}\,.
\end{equation}

\subsection{The ghost sector}
\label{sec:ghost}

The ghost sector of string field theory is more complicated than the matter
sector, due to the mid-point insertion required in the star product and
integration. The origin of this insertion is the singularity in the
worldsheet metric.
It is easiest to write the star-product in the bosonized ghost sector
as was done originally in~\cite{Witten:1986cc}.
It is most naturally defined in the half-string formulation, where
the mid-point is one of the basic degrees of freedom.

The mode expansion of the bosonized ghost $\phi(\s)$ is the same as the matter
coordinates with the exception that the zero-mode momentum is discretized
to half-integer values.
This momentum gives the ghost number of the state by $N_G=\hat p_0+3/2$,
where $\hat p_0$ is anti-Hermitian and it satisfies the following relations
with its eigenstate
\begin{equation}
\begin{aligned}
\hat p_0\ket{p_0}&=p_0\ket{p_0},&
\bra{p_0}\hat p_0&=-p_0\bra{p_0},\\
\braket{p_0}{p_0'}&=\delta_{p_0+p_0'}\,,\quad&
\sum_{p_0}\ket{p_0}\bra{-p_0}&=\One\,,\quad
\left(\ket{p_0}\right)^\dagger=\bra{p_0}\,.
\end{aligned}
\end{equation}

The star product and the integration in the ghost sector with the insertion
at the mid-point $\phi_m=\phi(\pi/2)$ can be written as
\begin{equation}
\begin{gathered}
(\Psi_1\star\Psi_2)(l^\ka,r^\ka,\phi_m)=e^{\frac{3i}{2}\phi_m}
    \int Dy^\ka \Psi_1(l^\ka,y^\ka,\phi_m)\Psi_2(-y^\ka,r^\ka,\phi_m)\,,\\
\int\Psi=\int Dy^\ka d\phi_m e^{-\frac{3i}{2}\phi_m}
    \Psi(y^\ka,-y^\ka,\phi_m)\,.
\end{gathered}
\end{equation}
To see that the integral vanishes unless $\Psi$ has ghost number 3,
we write a state with momentum $p_0$ as
$\braket{\phi_0}{\Psi,p_0}=\exp(ip_0\phi_0)\ket{\Psi}$.
The relation between $\phi_0$ and $\phi_m$ is as in~(\ref{MidPoint}),
implying that the integral vanishes unless $p_0=3/2$, i.e. $N_G=3$.
For the same reason the zero mode momentum obeys ${p_0}_3={p_0}_1+{p_0}_2+3/2$
under the star product and therefore the ghost number is additive.
This leads to the known facts that solutions of the projection equation
must have ghost number zero, and solutions of string field theory
must have ghost number one.

The vertices states in the continuous half-string basis are of the form
\begin{equation}
\label{vertices}
\ket{V_N}=\gamma_N
    \exp\left(-\frac{1}{2}{a_h^s}^\dagger {V_N^h}_{st}{a_h^t}^\dagger\right)
    \sum_{p_m^s}
    \delta_{\sum_s p_m^s-\frac{3}{2}(N-2)}\prod_{s=1}^N \ket{\Omega_h,p_m^s}\,.
\end{equation}
The mid-point momenta $p_m^s$ are the relevant degrees of freedom for the
half-string, but they are equal to the zero-mode momenta $p_m=p_0$.
The normalization factors $\gamma_N$ can be calculated using the fact that
the overlap of two surface states should be 1~\cite{LeClair:1989sj}.
We use the vacuum state, which in the continuous half-string basis is
\begin{equation}
\label{VacuumHalf}
\begin{aligned}
\ket{\Omega,p_0}&=N_\Omega e^{\mu_h a_h^\dagger}
    e^{\frac{1}{2}a_h^\dagger S a_h^\dagger}\ket{\Omega_h,p_m}\,,\\
\mu_h&=-\frac{1}{\sqrt{2}}(\One-S)J_\ka^h p_m\,,\qquad
N_\Omega=\det(\One-S^2)^{\frac{D+1}{4}}
    e^{-\frac{1}{2}\mu_h^T(1-S)^{-1}\mu_h}\,,
\end{aligned}
\end{equation}
where we used the transformation~(\ref{creationTrans}). The vacuum state
has a ghost number zero, meaning $p_0=-\threehalves$, but to calculate
overlaps we also need the shifted vacuum state with $p_0=\threehalves$.
The overlap of the vacuum with the identity (the one-vertex) is
\begin{equation}
\label{<0|I>}
\braket{\Omega,\threehalves}{\cI}=\gamma_1 N_\Omega
    \det(\One+CS)^{-\frac{D+1}{2}}
    e^{-\frac{1}{2}\mu_h^T(1-CS)^{-1}C\mu_h}\,.
\end{equation}
There are two sources of divergence in this expression, the determinants
and the exponents.
For the determinants, we use the relation $\det M=\exp(\tr\log M)$.
If the matrix $M$ is diagonal in the $\ka$ basis with eigenvalues
$m_\ka$, we have $\tr\log M=\int d\ka\,\rho(\ka)\log m_\ka$, where the
spectral density $\rho(\ka)$ diverges. In level truncation it diverges
as $\frac{\log L}{2\pi}$. 
In~\cite{Fuchs:2002wk,Belov:2002sq} it was calculated with its finite
contributions
\begin{equation}
\begin{aligned}
\rho^L(\ka)&=
    \frac{1}{2\pi}\sum_{n=1}^{L/2}\frac{1}{n}+\rho^L_{\text{fin}}(\ka)\,,\\
\rho_{\text{fin}}(\ka)&=
\lim_{L\rightarrow\infty}\rho^L_{\text{fin}}(\ka)=
\frac{4\log(2)-2\gamma-\Psi(\frac{i \ka}{2})-\Psi(-\frac{i \ka}{2})}{4\pi}\,.
\end{aligned}
\end{equation}
The divergence in the exponents comes from the scalar product in the $\ka$
basis, which results in integrals of $\frac{1}{\ka^2}$. This
divergence can also be regulated in level truncation by noticing that
\begin{equation}
\begin{aligned}
\lim_{L\rightarrow\infty}\sum_{n=1}^{L/2}\frac{1}{n}&=
\lim_{L\rightarrow\infty}
    \sum_{n,m=1}^L \frac{J_n J_m}{\sqrt{nm}}\delta_{nm}\\
&=\int_{-\infty}^\infty d\ka
    \sum_{n=1}^\infty\frac{J_n v_n^\ka}{\sqrt{n}}
    \sum_{m=1}^\infty\frac{J_m v_m^\ka}{\sqrt{m}}=
\int_{-\infty}^\infty d\ka\,J_\ka J_\ka=
\int_0^\infty {\cal P}\frac{4d\ka}{\ka^2\N(\ka)}\,.
\end{aligned}
\end{equation}
Therefore, the two types of divergences can be compared.

We now have everything we need to calculate $\gamma_1$ from the requirement
that the overlap~(\ref{<0|I>}) equals 1
\begin{equation}
\gamma_1=e^{-(D+1)(\frac{1}{8}\sum\frac{1}{n}+d_1)}
    e^{\frac{9}{8}\sum\frac{1}{n}}\,,\qquad
d_1=\frac{1}{4}\int_0^\infty d\ka\,\rho_{\text{fin}}(\ka)
    \log\left(\coth^2\frac{\ka\pi}{4}\right)=\frac{\log(2\pi)-\gamma}{8}\,,
\end{equation}
where the first exponent comes from the determinant and the second is
the exponent from the ghost sector. The integral of $d_1$ was calculated
using integration by parts and the integral representation of $\log\Gamma$.
We can calculate the normalizations of all the vertices using the relations
\begin{equation}
\label{gammaN}
\ket{V_N}=\braket{\cI}{V_{N+1}} \Rightarrow \gamma_N=\gamma_1\gamma_{N+1}
    \Rightarrow \gamma_N=\gamma_1^{2-N}\,.
\end{equation}

Next, we do a couple of calculations, to check the validity of our results.
Repeating the calculation for the reflector (the two-vertex) gives
\begin{equation}
\vphantom{\bra{\threehalves}}_1\!
\bra{\Omega,\threehalves}
\vphantom{\bra{\threehalves}}_2\!
\braket{\Omega,-\threehalves}{\cal R}_{12}=
    \gamma_2 N_\Omega^2
    \det(\One+V^h_2 S_2)^{-\frac{D+1}{2}}
    e^{-\frac{1}{2}\mu_{2h}^T(1-S_2 V^h_2)^{-1}V^h_2\mu_{2h}}=1\,,
\end{equation}
where $S_2=\One_2\otimes S, \mu_{2h}=(\mu_h^1,\mu_h^2)$.
Here we get $\gamma_2=1$, independently of the number of dimensions $D$,
in accord with~(\ref{gammaN}). This result is analogous
to the calculation of the overlap of the vacuum state with itself
\begin{equation}
\ket{\Omega,p_0}=\braket{\Omega,p_0}{\cal R}\Rightarrow
\braket{\Omega,\threehalves}{\Omega,-\threehalves}=
    \bra{\Omega,\threehalves}\braket{\Omega,-\threehalves}{\cal R}=1\,.
\end{equation}
Next, we calculate explicitly the normalization of the three-vertex
\begin{equation}
\label{<0|V3>}
\begin{aligned}
\vphantom{\bra{\threehalves}}_1\!
\bra{\Omega,\threehalves}&
\vphantom{\bra{\threehalves}}_2\!
\bra{\Omega,-\threehalves}
\vphantom{\bra{\threehalves}}_3\!
    \braket{\Omega,-\threehalves}{V_3}_{123}=
    \gamma_3 N_\Omega^3
    \det(\One+V^h_3 S_3)^{-\frac{D+1}{2}}
    e^{-\frac{1}{2}\mu_{3h}^T(1-S_3 V^h_3)^{-1}V^h_3\mu_{3h}}=1\\
\Rightarrow \gamma_3&=e^{(D+1)(\frac{5}{72}\sum\frac{1}{n}+d_3)}
    e^{\frac{3}{8}\sum\frac{1}{n}+e_3}\,,\qquad
e_3=\frac{3}{2}\log\frac{27}{16}\,,\\
d_3&=-\frac{1}{4}\int_0^\infty d\ka\,\rho_{\text{fin}}(\ka)
\log\left(4\csch(\frac{3\ka\pi}{4})^4 \sinh(\frac{\ka\pi}{2})^6\right)\\
   &=\frac{1}{72}\left(25\log2-11\log3-3\log\pi-5\gamma
      +24\log\Gamma(\frac{1}{3})+96\zeta'(-1)\right)\approx 0.09296\,.
\end{aligned}
\end{equation}
Comparing the diverging terms, we get that the relation
$\gamma_3=\gamma_1^{-1}$ holds only for the critical dimension $D=26$.
Unfortunately, the finite parts do not match.
Such a discrepancy also appeared in the calculation of the overlap of
the two wedge states $\braket{3}{3}$ in~\cite{Belov:2002pd}.
This means that the level truncation regularization we are using does not
preserve the Virasoro algebra.
This anomaly can come from the $\kappa=0$ state which we ignored or from
the calculation of the determinants.
It can also come from using the bosonized ghost.
Truncating the fermionic ghost $c=:e^{i\phi}:$ to level $L$ is not the same
as truncating the bosonized ghost $\phi$ to the same level, which can be
seen from their relation $(c)_L=(:e^{i\phi}:)_L\neq :e^{i(\phi)_L}:$.

We conclude this section by calculating the full-string three-vertex in
the bosonized ghost sector from its definition in the half-string
basis~(\ref{vertices}). For this we need to use the transformation
\begin{equation}
a_\ka^\dagger=W a_h^\dagger-U a_h+\frac{1}{\sqrt{2}}J_\ka p_m\,.
\end{equation}
Using again the results of~\ref{sec:TofStates} we get
\begin{equation}
\ket{V_3(P_0)}=\gamma_3\det\!\left(\frac{\One-V_3^2}{\One-{V_3^h}^2}\right)^{1/4}
\!\!\!\!\!\!\delta_{p_0^1+p_0^2+p_0^3-\threehalves}
    \exp\!\left(-\frac{1}{2}A^\dagger V_3 A^\dagger
    +P_0 V_0 A^\dagger-\frac{1}{2}P_0 V_{00}P_0\right)\!\ket{\Omega}_{123} .
\end{equation}
The vertex is different from the three-vertex of the matter sector,
requiring the following redefinition of the three-vertex matrices
\begin{equation}
\begin{aligned}
V_3^{st}&\rightarrow V_3^{st}\,,\\
V_0^{st}&\rightarrow V_0^{st}+\frac{\sqrt{2}}{3}J_\ka\,,\\
V_{00}^{st}&\rightarrow V_{00}\delta^{st}+\frac{1}{3}\sum\frac{1}{n}
    -\frac{1}{6}\log\frac{27}{16}\,.
\end{aligned}
\end{equation}
The vertex of the matter sector~(\ref{V3full}) is recovered if we set
the momenta conservation relation $p_0^1+p_0^2+p_0^3=0$.
The determinant factor cannot be calculate directly, since it is zero
divided by zero. We calculate it using the relation between $V_3$ and
$V_3^h$
\begin{equation}
V_3=(W_3+V_3^h U_3)^{-1}(V_3^h W_3+U_3)\Rightarrow
\det\left(\frac{\One-V_3^2}{\One-{V_3^h}^2}\right)=
\det\left(\frac{\One-S_3^{\,2}}{(\One+V_3^h S_3)^2}\right)\,.
\end{equation}
As a check of the transformation, we repeat the calculation~(\ref{<0|V3>})
in the full-string basis
\begin{equation}
\begin{aligned}
\vphantom{\bra{\threehalves}}_1\!
\bra{\Omega,\threehalves}
\vphantom{\bra{\threehalves}}_2\!
\bra{\Omega,-\threehalves}
\vphantom{\bra{\threehalves}}_3\!
    \braket{\Omega,-\threehalves}{V_3}_{123}=
    \gamma_3
    \det\left(\frac{\One-S_3^{\,2}}{(\One+V_3^h S_3)^2}\right)^{\frac{D+1}{4}}
    e^{-\frac{1}{2}P_0 V_{00} P_0}=1\,.
\end{aligned}
\end{equation}
This demonstrates that the anomaly we got is not related to the
transformation. It is related either to the $\ka$ basis or to the
bosonized ghost or to both.

\section{Applications}
\label{sec:app}

Now that we have the continuous half-string formalism, we can use it in various
calculations.
The multiplication of two $\HH_{\ka^2}$ states becomes very simple in
this representation. Parameterizing two states $\cS_i$ by
\begin{equation}
S^h_i=\mat{a_i & b_i \\ b_i & c_i}\,,
\end{equation}
where $a_i,b_i,c_i$ are functions of $\ka>0$,
we get for $\cS^h\equiv \cS^h_1\star\cS^h_2$,
\begin{equation}
\label{HSMult}
S^h=\frac{1}{1-a_2 c_1}\mat{a_1(1-a_2 c_1)+a_2 b_1^2
 & -b_1 b_2 \\  -b_1 b_2 & c_2(1-a_2 c_1)+c_1 b_2^2}\,.
\end{equation}
This expression is equivalent to a gaussian integration in the functional
language (up to normalization)
\begin{equation}
\braket{l^\ka,r^\ka}{\cS^h}
=
\braket{l^\ka,r^\ka}
{\cS^h_1\star \cS^h_2}
=\int dy_\ka 
  e^{-\frac{1}{2}\mat{l^\ka&y^\ka}{\displaystyle L^h_1}\mat{l^\ka\\y^\ka}}
  e^{-\frac{1}{2}\mat{-y^\ka&r^\ka}{\displaystyle L^h_2}\mat{-y^\ka\\r^\ka}}\,,
\end{equation}
where
\begin{equation}
L^h=\frac{\One-S^h}{\One+S^h}\,.
\end{equation}

\subsection{Doubling of the wedge subalgebra}
\label{sec:doubling}

The wedge states $\ket{n}$ are squeezed states, given by the
matrix~\cite{Furuuchi:2001df}
\begin{equation}
\Sigma_n=C T_n\,,
\end{equation}
where
\begin{equation}
T_n=\frac{T+(-T)^{n-1}}{1-(-T)^n}\,,\qquad
T=-e^{-\frac{\ka \pi}{2}}\mat{1 & 0 \\ 0 & 1}\,.
\end{equation}
The case $n=1$ gives the identity state $\ket{\cI}$,
$n=2$ gives the vacuum state $\ket{\Omega}$ and
$n\rightarrow\infty$ is the sliver $\ket{S}$.

Using~(\ref{SofVUW}), we get the expression for the wedge matrices
$\Sigma_n$ in the sliver basis
\begin{equation}
\label{wedgeS}
\Sigma^h_n=-e^{-\frac{\ka \pi}{2}(n-1)}\mat{0 & 1 \\ 1 & 0}\,.
\end{equation}
The case $n\rightarrow \infty$ gives
\begin{equation}
\Sigma^h_\infty=\mat{0 & 0 \\ 0 & 0}\,,
\end{equation}
which is the sliver matrix in the sliver basis.
The case $n=2$ gives
\begin{equation}
\Sigma^h_2=-\Sigma_\infty\,,
\end{equation}
meaning that the matrix defining the vacuum state in the sliver basis
is minus the sliver matrix in the regular basis, as expected~(\ref{InvRule}).

Using the expression for the star product of two $\HH_{\ka^2}$
states in the sliver basis~(\ref{HSMult}), we verify the known
algebra
\begin{equation}
\label{wedgeAlgebra}
\ket{n}\star\ket{m}=\ket{n+m-1}\,.
\end{equation}
By inspecting eq.~(\ref{HSMult}),(\ref{wedgeS}), we recognize that this algebra
can be doubled by defining $\Sigma_n^+=\Sigma^h_n$ and
$\Sigma_n^-=-\Sigma^h_n$. The corresponding states $\ket{n,\pm}$ form
a subalgebra of the star product with the product given by
\begin{equation}
\ket{n,r}\star\ket{m,s}=\ket{n+m-1,rs}\,,
\end{equation}
where $r,s=\pm 1$.
This subalgebra is a direct product of ${\mathbb Z}_2$ with the wedge
subalgebra, up to the identification of the sliver
$\ket{\infty,+}=\ket{\infty,-}$.

\subsection{The gauge transformation of the sliver to other butterflies}
\label{sec:gauge}

The spectroscopy of the general butterflies performed in~\cite{Fuchs:2002zz}
supported the natural assumption that in vacuum string field theory
all these states are related by gauge transformations.
It is possible to define states with different matter and ghost content.
In vacuum string field theory it is common to assume a universal ghost
part. Moreover, it was shown in~\cite{Fuchs:2002zz}
that states with the same ghost factor, whose matter factor consists of
different matter butterflies, are orthogonal.
In this subsection we explicitly present the gauge transformations among
such states, using the gauge factorization ansatz of~\cite{Imamura:2002rn}.
Therefore, we treat only the matter sector.
It should be possible to perform similar gauge transformation
in the ghost sector due to the enlarged gauge freedom of vacuum string field
theory~\cite{Gaiotto:2001ji}.

The generalized butterflies can be described by the parameter
$0\leq a \leq \infty$, which is related to $\alpha$ of~\cite{Gaiotto:2002kf}
by $a=\frac{2-\alpha}{\alpha}$. According to the butterfly spectroscopy,
they are in $\HH_{\ka^2}$ and their defining matrices were found to be
\begin{equation}
S_a=\frac{1}{\sinh(\frac{\ka\pi}{2}(1+a))}
  \mat{\sinh(\frac{\ka\pi}{2}) & \sinh(\frac{\ka\pi}{2} a) \\
       \sinh(\frac{\ka\pi}{2} a) & \sinh(\frac{\ka\pi}{2})}\,,
\end{equation}
where $a=0$ is the nothing state, $a=1$ is the canonical butterfly,
and the limit $a \rightarrow \infty$ is the sliver.
Using these results and eq.~(\ref{SofVUW}), we find that in the sliver basis
they transform to
\begin{equation}
\label{GBinSliver}
S^h_a=e^{-\frac{\ka\pi}{2} a}\mat{1 & 0 \\
                 0 & 1}\,.
\end{equation}
The left--right factorization of the projectors is evident from the
structure of the matrix. The off diagonal entries, which mix left
and right modes, vanish.
If we take a general two-dimensional matrix,
the condition that its off-diagonal entries vanish in the sliver basis
is (once again) the projection condition of~\cite{Fuchs:2002zz}.

An infinitesimal gauge transformation is given by
\begin{equation}
\label{infGauge}
\delta \Phi=\Phi \star \Lambda-\Lambda \star \Phi\,.
\end{equation}
Exponentiating this transformation, we get
\begin{equation}
\label{fullGT}
\Phi \rightarrow \cU^{-1}\star\Phi \star \cU\,,
\end{equation}
where
\begin{equation}
\cU=\exp_\star(\Lambda)=\cI+\Lambda+\frac{1}{2}\Lambda\star\Lambda+...\,.
\end{equation}
Here, $\cI$ is the star-algebra identity element and $\cU^{-1}$
is the inverse with respect to the star product of $\cU$, i.e.
\begin{equation}
\cU\star \cU^{-1}=\cU^{-1}\star \cU=\cI\,.
\end{equation}

For simplicity, we limit the search to squeezed states in $\HH_{\ka^2}$
whose defining matrix is real.
Given a squeezed state $\cU$ in $\HH_{\ka^2}$, defined by the matrix
$U^h$, we find using~(\ref{HSMult})
that $\cU^{-1}$ is also a squeezed state given by $C {U^h}^{-1}C$.
It can be inferred
from~\cite{Bars:2002nu} that this result is not limited to $\HH_{\ka^2}$.
We shall not restrict ourselves to BPZ-real states, 
since $\cU$ is not a physical state, but a gauge transformation.

For a real matrix to define a normalizable squeezed state all its eigenvalues
$\lambda$ should be less than unity in absolute value.
As we allow for singular squeezed states, such as the identity,
we relax the condition to $|\lambda| \leq 1$.
Define $\lambda_{1,2}$ to be the eigenvalues of $U^h$.
The eigenvalues of ${U^h}^{-1}$ are $\frac{1}{\lambda_{1,2}}$. The
conditions
\begin{equation}
|\lambda_{1,2}| \leq 1\,,\,\,|\frac{1}{\lambda_{1,2}}| \leq 1\,,
\end{equation}
imply
\begin{equation}
|\lambda_{1,2}| = 1\,.
\end{equation}
As we assumed that $U^h$ is real and, being a quadratic form, it is
necessarily symmetric, it follows that both eigenvalues are real, and thus
\begin{equation}
\lambda_{1,2} = \pm 1\,.
\end{equation}
We would like to find a one-parameter family $U_a$ which
relates the sliver to any other generalized butterfly.
Continuity with respect to $a$, and $U_\infty=C$, imply
that one eigenvalue equals $+1$, and the other equals $-1$.
To summarize, $U^h$ is real, symmetric, traceless and $\det(U^h)=-1$.

A natural parameterization is
\begin{equation}
\label{paramTheta}
U(\theta)=\mat{\sin(\theta)  &  -\cos(\theta)  \\
        -\cos(\theta)  &  -\sin(\theta)}\,,
\end{equation}
where $\theta=0$ is the identity state.
With this parameterization the multiplication rule~(\ref{HSMult})
translates into
\begin{equation}
\label{badParam}
\cU(\theta)\star\cU(\phi)=\cU(\psi)\,,\qquad
   \sin(\psi)=\frac{\sin(\theta)+\sin(\phi)}{1+\sin(\theta)\sin(\phi)}\,,
\end{equation}
so indeed we get a one parameter group.
This multiplication rule is singular around $\sin(\theta)\sin(\phi)=-1$.
Although $\sin(\theta)\sin(\phi)\neq -1$ implies
$|\frac{\sin(\theta)+\sin(\phi)}{1+\sin(\theta)\sin(\phi)}|\leq 1$
and so the multiplication rule is well defined in a neighborhood of
the singular points, the singularity is real, as the limit of
the expression at the singular points is direction dependent.
The multiplication $\cU(\frac{\pi}{2})\star \cU(-\frac{\pi}{2})$ can be
regularized in different ways so as to produce any $\cU(\theta)$ as a result.
In string field theory we got used to anomalies of the star product at
$\ka=0$. However, here the anomaly can appear at any $\ka$.

The group of gauge transformations (\ref{paramTheta}) with
$-\frac{\pi}{2}<\theta<\frac{3\pi}{2},\theta\neq\frac{\pi}{2}$ is a direct product of
${\mathbb Z}_2$ with the subgroup parameterized by
$-\frac{\pi}{2}<\theta<\frac{\pi}{2}$. The elements of the ${\mathbb Z}_2$
are the identity $\cI=\cU(0)$ and the identity dual $\cJ=\cU(\pi)$.
The two topologically disconnected sections of the gauge group are
related by $\cU(\pi-\theta)=\cJ\star\cU(\theta)$.
$\cJ$ behaves as a root of the identity $\cJ\star\cJ=\cI$, and it
transforms all squeezed states to themselves
\begin{equation}
\cS=\cJ^{-1}\star\cS\star\cJ\,,
\end{equation}
or in other words, it star-commutes with all squeezed states.
This can be seen most easily in the continuous half-string functional form
\begin{equation}
\braket{l^\ka,r^\ka}{J}=\delta(l^\ka-r^\ka)\Rightarrow
\begin{cases}
\Psi[l^\ka,r^\ka]\star\cJ=\Psi[l^\ka,-r^\ka]\\
\cJ\star\Psi[l^\ka,r^\ka]=\Psi[-l^\ka,r^\ka]
\end{cases}.
\end{equation}
Therefore, $\cJ$ star-commutes with all states in which $l^\ka,r^\ka$
appear in even powers, and particularly with all squeezed states,
meaning that these states are not charged under the ${\mathbb Z}_2$.

The subgroup parameterized by $-\frac{\pi}{2}<\theta<\frac{\pi}{2}$,
can be parametrized as
\begin{equation}
\label{goodParam}
U(t)=\mat{\tanh(t)  &  -\sech(t)  \\
        -\sech(t)  &  -\tanh(t)}\,,
\end{equation}
where $-\infty<t<\infty$. Now the star-multiplication rule simplifies to
a linear relation
\begin{equation}
\label{linearGauge}
\cU(t_1)\star\cU(t_2)=\cU(t_1+t_2)\,,
\end{equation}
and the anomaly is hidden at $t=\pm\infty$.
In this parameterization all gauge transformations are generated by
infinitesimal gauge transformations $\Lambda$.

We can calculate $\Lambda$~(\ref{infGauge}) as was done
in~\cite{Imamura:2002rn}. Temporarily restoring the $\ka$ dependence, we have
\begin{equation}
\ket{\cU[t_\ka]}=\exp\left(-\frac{1}{2}\int_0^\infty d\ka 
  \mat{a^{\dagger\,\ka}_l & a^{\dagger\,\ka}_r}
   \mat{\tanh(t_\ka)  &  -\sech(t_\ka)  \\
        -\sech(t_\ka)  &  -\tanh(t_\ka)}
   \mat{a^{\dagger\,\ka}_l \\ a^{\dagger\,\ka}_r}\right) \ket{\Omega_h},
\end{equation}
from which we get
\begin{equation}
\ket{\Lambda[t_\ka]}=\left.\frac{d}{ds}\ket{\cU[st_\ka]}\right |_{s=0}=
 -\frac{1}{2}\int_0^\infty d\ka\,t_\ka
 (a^{\dagger\,\ka}_l a^{\dagger\,\ka}_l 
 -a^{\dagger\,\ka}_r a^{\dagger\,\ka}_r) \ket{\cI}.
\end{equation}

For finding a gauge transformation $\cU_a$ from the sliver to a generalized
butterfly~(\ref{GBinSliver}), we write
\begin{equation}
\cU_a^{-1}\star \cS^h_\infty \star \cU_a=\cS^h_a\,,
\end{equation}
and use the expression~(\ref{goodParam}) to get
\begin{equation}
\tanh(t)=-e^{-\frac{\ka \pi}{2}a}\,,\qquad
  U_a=\mat{-e^{-\frac{\ka \pi}{2}a} & -\sqrt{1-e^{-\ka \pi a}} \\
                 -\sqrt{1-e^{-\ka \pi a}} & e^{-\frac{\ka \pi}{2}a}}\,.
\end{equation}
We see that the gauge transformation approaches the singular limit in all cases
when $\ka \rightarrow 0$. All the butterflies share the eigenvalue $1$ for
$\ka=0$, and thus should have a trivial gauge transformation for this value.
The discontinuity of the transformation near $\ka=0$ is related to the
discontinuity of the transformation to the sliver basis.
For the nothing state $a=0$, the transformation is singular for all $\ka$,
due to the singular eigenvalues of the nothing.
Thus, the nothing should not be thought of as equivalent to the other states.
The singularity of the nothing can also be seen if we try to find a
Bogoliubov transformation to the nothing basis. This transformation
is singular and has to be regularized, and the final result of the
transformation is regularization dependent.

Composing two gauge transformations, we can define the transformation
$\cU_{ab}$ between two butterflies
\begin{equation}
\cU_{ab}^{-1}\star \cS^h_b \star \cU_{ab}=\cS^h_a\,,
\end{equation}
using
\begin{equation}
\begin{aligned}
\tanh(t)&
    =\frac{\sinh(\frac{\ka \pi}{4}(a-b))}{\sinh(\frac{\ka \pi}{4}(a+b))}\,,\\
U_{ab}&=\frac{1}{\sinh(\frac{\ka \pi}{4}(a+b))}\mat{
   \sinh(\frac{\ka \pi}{4}(a-b)) &
                 -\sqrt{\sinh(\frac{\ka \pi}{2}a) \sinh(\frac{\ka \pi}{2}b)}\\
   -\sqrt{\sinh(\frac{\ka \pi}{2}a)\sinh(\frac{\ka \pi}{2}b)} &
                 -\sinh(\frac{\ka \pi}{4}(a-b))}\,.
\end{aligned}
\end{equation}

It is instructive to imply the gauge transformation $\cU_1$ to the wedge
states. This transformation sends the sliver to the butterfly, and so should
transform the wedge states to ``butterfly wedge states''. These states form
a subalgebra isomorphic to that of wedge states, and can be used as a
regularization of the butterfly. 
Using $U_1$ and the expression~(\ref{wedgeS}) for the wedge states,
we get that the butterfly wedge states in the sliver basis are
defined by the matrix
\begin{equation}
{\Sigma^b_n}^h=\frac{1}{\sinh(\frac{\ka \pi}{2}n)}
        \mat{\sinh(\frac{\ka \pi}{2}(n-1)) & -\sinh(\frac{\ka \pi}{2}) \\
         -\sinh(\frac{\ka \pi}{2}) & \sinh(\frac{\ka \pi}{2}(n-1))}\,.
\end{equation}
In particular we see that the identity state $n=1$ is invariant, while
the sliver $n=\infty$ transforms into the butterfly.
This construction could have been especially useful if the butterfly wedge
states were surface states.
This could have been the case if $\cU_1,\,\cU_1^{-1}$ were surface states.
We can see that this is not the case. The vacuum state $n=2$ transforms to
${\Sigma^b_2}^h$, which in the full-string $\ka$ basis becomes
\begin{equation}
\Sigma^b_2=\mat{e^{-\frac{\ka \pi}{2}} & 0 \\
        0 & e^{-\frac{\ka \pi}{2}}}\,.
\end{equation}
Using the methods of section $3$ of~\cite{Fuchs:2002zz} we find that the
candidate conformal transformation of this state is $\tan^{-1}(z)$, which
describes the sliver. Thus, $\Sigma^b_2$ is not a surface state.

\subsection{The tachyon}
\label{sec:tachyon}

We start by making the simplest calculation of the three-point function
of the tachyon. This calculation is much simpler in the full-string basis,
but it demonstrates that such calculations can also be made in the
continuous half-string basis.

The tachyon state with momentum $p$ is simply the vacuum state with shifted
momentum $\ket{T(p)}=\ket{\Omega,p}$.
To calculate the tachyon's three-point function, we need to choose a basis.
The tachyon functional in the sliver half-string basis is
\begin{equation}
\begin{aligned}
T(l^\ka,r^\ka,p)&=\int dx_m e^{ip_0x_0}
    e^{-\frac{1}{2}\int d\ka x^\ka_h L_\ka x^\ka_h}\,,\qquad&
x_0&=x_m-\int d\ka J^h_\ka x^\ka_h\,,\\
L_\ka&=\frac{1}{\sinh(\frac{\ka\pi}{2})}
    \mat{\cosh(\frac{\ka\pi}{2})&1\\1&\cosh(\frac{\ka\pi}{2})},\qquad&
J^h_\ka&={\cal P}
\frac{1}{\sqrt{2\ka}\cosh(\frac{\ka\pi}{4})}\mat{-1\\1}.
\end{aligned}
\end{equation}
The calculation involves gaussian
integrations over an infinite matrix, but this matrix is diagonal.
Tracing the factors of $\ka$, we get (up to normalization)
the known result
\begin{equation}
\left<T\star T\star T\right>=
\exp(-\frac{(2\alpha')}{4}(p_1^2+p_2^2+p_3^2)\log\frac{27}{16})\,.
\end{equation}
It would be most interesting to calculate the four-tachyon amplitude, and
to reproduce the Veneziano amplitude. This calculation is much more
complicated, because it involves the propagator, which is fairly
complicated in the half-string basis.

Next, we
discuss the results of~\cite{Hata:2001sq} that found
the tachyon solution around the D-brane of VSFT.
Their calculation gave the correct tachyon
mass~\cite{Rastelli:2001wk,Hata:2002it}, but resulted in a wrong
value for the D-brane tension.
Later, the tachyon solution was calculated in a regularized way
in~\cite{Okawa:2002pd} giving the correct tension.

Hata and Kawano considered the D-brane solution
\begin{equation}
\label{SliverEq}
\Phi_c\star\Phi_c+\Phi_c=0\,,
\end{equation}
where $-\ket{\Phi_c}=\ket{S}$ is the sliver.
They looked for a solution of
the linearized equation of motion around
$\ket{\Phi_c}$
\begin{equation}
\label{teq}
\Phi_t+\Phi_t\star\Phi_c+\Phi_c\star\Phi_t=0\,,
\end{equation}
where $\Phi_t$
represents the tachyon around the D-brane.
They implicitly assumed that $\Phi_t$ star-commutes with $\Phi_c$.
Thus, eq.~(\ref{teq}) reduces to
\begin{equation}
\label{TEOM}
\Phi_c\star\Phi_t=\Phi_t\star\Phi_c=-\frac{1}{2}\Phi_t\,.
\end{equation}
Formally there is no non-trivial solution to this equation.
Multiplying this equation by $\Phi_c$ from the left (right),
and using~(\ref{SliverEq}) gives
\begin{equation}
\Phi_c\star\Phi_t=0\,, \qquad \Phi_t\star\Phi_c=0\,,
\end{equation}
and thus
\begin{equation}
\Phi_t=0\,.
\end{equation}
However, the HK state
\begin{equation}
\ket{\Phi_t}=e^{-t_n a_n^\dagger p_0+ip_0x_0}\ket{\Phi_c},
\end{equation}
inherits the singular properties of the sliver.
Both $\Phi_c$ and $\Phi_t$ should be regularized.
The tachyon state can satisfy its equations of motion in a
weak sense, provided
the regularization scheme does not respect the projection
equation~(\ref{SliverEq}).
Indeed, both the level-truncation regularization
of~\cite{Hata:2001sq,Hata:2001wa,Hata:2002it}
and the wedge state regularization of~\cite{Rastelli:2001wk},
satisfy the projection equation only at the limit.

We now transform the tachyon state to the continuous half-string
basis. This is done in the same manner we transformed the three-vertex.
We use the transformation~(\ref{creationTrans})
and the results of~\ref{sec:TofStates} to write
\begin{equation}
\label{tBogo}
\ket{\Phi_t}=
    e^{t_\ka a_\ka^\dagger p_0}e^{ip_0 x_0}\ket{\Phi_c}=
-e^{\frac{1}{2}t_\ka^T(1+S)^{-1}t_\ka p_m^2-\frac{1}{2}{t_\ka^h}^T(1+S)^{-1}t_\ka^h p_m^2}
    e^{t_\ka^h {a_\ka^h}^\dagger p_m}e^{ip_m x_m}\ket{\Omega_h},
\end{equation}
where $t$ is given in the $\kappa$ basis by
\begin{equation}
t_\ka=3(1+CS)(1+3CV^{11})^{-1}V_0^{11}=
    \frac{\sqrt{\N(\kappa)}}{1+e^{\frac{\ka\pi}{2}}}\mat{-1\\1}.
\end{equation}
Notice that $t_\ka$ is an odd function of $\ka$ as
required by twist invariance, but it
has a discontinuity at $\ka=0$.
The linear term of the tachyon state in the half-string
basis~(\ref{tBogo}) vanishes
\begin{equation}
t^h_\ka=0\,,
\end{equation}
and this gives another expression for $t_\ka$,
\begin{equation}
t_\ka=\frac{1}{\sqrt{2}}W^{-1}J^h_\ka=\frac{1}{\sqrt{2}}(1+S)J_\ka\,.
\end{equation}
The normalization factor of $\Phi_t$ does not affect its equation of
motion, since~(\ref{teq}) is linear in $\Phi_t$. It is still worth noting
that the $p_m^2$ coefficient $\frac{1}{2}t^T(1+S)^{-1}t$ diverges.
Up to the normalization factor, the tachyon state is given by the
new vacuum with a mid-point momentum insertion
\begin{equation}
\ket{\Phi_t}=
e^{ip_m x_m}\ket{\Omega_h},
\end{equation}
as can be inferred from the geometric construction
of~\cite{Rastelli:2001wk}.
This state cannot satisfy~(\ref{TEOM}),
in accord with the discussion above.
Therefore, a regularization of this calculation
in the continuous half-string basis is needed.

We regularize the tachyon and sliver states using the wedge states $\ket{n}$.
In the half-string basis we define the wedge states in the matter sector so
as to satisfy~(\ref{wedgeAlgebra}) and the three-vertex as
a squeezed state with no other normalization as in~(\ref{V3matter}).
In this convention the overlap of two wedge states is
\begin{align}
\begin{aligned}
N_{n,m} &\equiv \frac{1}{V}\braket{n}{m}=\det(\One-\Sigma_n^h\Sigma_m^h)^{-\frac{D}{2}}
& =\exp\left(\frac{D}{6(n+m-2)}\sum_{l=1}^{L/2}\frac{1}{l}+ \text{finite term}
\right) ,
\end{aligned}
\end{align}
where we evaluate the determinant as in subsection~\ref{sec:ghost}.
The finite term comes from $\rho_\text{fin}$ and vanishes
in the regularization limit.
The volume factor $V=\delta(0)$, comes from the two-vertex delta
function $\delta(p_1+p_2)$.
The regularized tachyon state is taken to be the wedge state with a mid-point
momentum insertion
\begin{align}
\ket{T_n(p_m)}=\ket{n,p_m}.
\end{align}

Following~\cite{Rastelli:2001wk} we check the normalization of the HK state
by taking the overlap of~(\ref{teq})
with the $p_0$ vacuum state~(\ref{VacuumHalf}).
The first term of~(\ref{teq}) gives
\begin{align}
\braket{\Omega,p_0}{T_n(p_m)}&=N_\Omega N_{2,n}
    e^{-\frac{1}{2}\mu^h(\One+\Sigma_n^h S)^{-1}\Sigma_n^h\mu^h}
    \delta(p_0+p_m)\,,
\end{align}
where we used the fact that
the vacuum state is proportional to the wedge state $\ket{2}$.
The second and third terms of~(\ref{teq}) give the same
state~\footnote{
Notice that this state is twist invariant even for finite $n$, which
demonstrates that the regularization we are using is
different from that of~\cite{Rastelli:2001wk} who
got twist invariant states only in the limit $n\rightarrow\infty$.
}
\begin{align}
\label{TnStar}
\ket{T_n(p_m)} \star\ket{n}=\ket{n} \star\ket{T_n(p_m)}=\ket{T_{2n-1}(p_m)}.
\end{align}
The ratio of the two overlaps is
\begin{align}
\frac
{\braket{\Omega,p_0}{T_{2n-1}(p_m)}}
    {\braket{\Omega,p_0}{T_n(p_m)}} &=
    \frac{N_{2,2n-1}}{N_{2,n}}
    e^{-\frac{1}{2}\mu^h(\One+\Sigma_{2n-1}^h S)^{-1}\Sigma_{2n-1}^h\mu^h}
    e^{\frac{1}{2}\mu^h(\One+\Sigma_n^h S)^{-1}\Sigma_n^h\mu^h}\,.
\end{align}
The expression in the exponent reads
\begin{align}
\label{ExpNorm}
\int_0^\infty d\ka
{\cal P} \frac{p_0^2}{2\ka}
\left(\coth(\frac{n\ka\pi}{4}) - \coth(\frac{(2n-1)\ka\pi}{4}) \right)
=-\frac{p_0^2}{2}\left(
\log(2)+O(n^{-1})
\right),
\end{align}
where we removed the $\frac{1}{\ka^2}$ singularity, which can be done
consistently due to the principal value, and the scheme we use is
not important because it only gives $O(n^{-1})$ contributions.
In the limit $n\rightarrow\infty$ we get
\begin{align}
\braket{\Omega,p_0}{\Phi_c\star \Phi_t}=e^{-\frac{p_0^2}{2}\log2}\braket{\Omega,p_0}{\Phi_t}\,,
\end{align}
reproducing the tachyon's equation of motion~(\ref{TEOM}) for $p_0^2=2$,
which is the mass of the tachyon in our convention, $2\alpha'=1$.

\section{Conclusions}
\label{sec:conclusions}

In this paper, we demonstrated how the transformation to the half-string
basis works in the continuous basis. The star-product in this formalism
is as simple as it gets. This allows us to show that all the butterfly
projectors, which are all supposed to describe the same D-brane, are
indeed related by a gauge transformation. In the half-string formalism
we were able to find a parameterization of the gauge
transformation~(\ref{goodParam}), which is linear under the
star-multiplication~(\ref{linearGauge}), and to write explicitly
the gauge transformation between the butterflies.

To compute string scattering amplitudes, we still need to refer to the
full string basis, in which the physical degrees of freedom are defined.
This was demonstrated in the calculation of the three tachyon amplitude.
It would be interesting to attempt to calculate the four tachyon
amplitude in this formalism.

Using the known form of the vertex in the half-string basis, we were
able to reconstruct the zero-mode dependence of the vertex in the
full-string basis, and the ghost sector. We used our knowledge on the
overlap of surface states to calculate the normalization of the
vertices in the continuous half string basis. These normalizations
are infinite. Yet, we can work out the relations between these
infinities in the level truncation regularization and get the correct
relations at the critical dimension $D=26$. The finite terms in
the normalizations do not give the right relations. This means that
the way we employ the level truncation does not preserve the
Virasoro algebra. This could be inherent to the continuous basis
or related to the use of the bosonic ghost sector.

It would be interesting to check the continuous basis by repeating
the calculations
of these normalizations
numerically in the discrete basis and compare
the results. If one can demonstrate that they differ, it would be
even more interesting to find a way to regularize the continuous
basis, perhaps by handling the $\ka=0$ eigenstate more carefully.

Our results can be generalized to the fermionic ghost.
To transform the fermionic ghost to the continuous half-string basis,
we need to repeat the analysis of the Bogoliubov transformation that
we presented in the appendix for anti-commuting operators. This will
also allow for generalizing our results to the superstring vertices.

We hope that the progress we made will help in the search for an analytic
solution of string field theory, as well as the search for analytic expressions
for scattering amplitudes~\cite{Taylor:2002bq,Coletti:2003ai,Bars:2002qt}.

\section*{Acknowledgments}

We would like to thank Ofer Aharony, Itzhak Bars, Dmitri Belov,
Theodore Erler, Yaron Oz, Ronen Segev,
Jacob Sonnenschein and Barton Zwiebach, for discussions.
M.~K. would like to thank Sergei Kuzenko and Ian McArthur from the
University of Western Australia, and Peter Bouwknegt
and Mathai Varghese from the University of Adelaide, for hospitality,
and Elias Kiritsis for the 
organization of the interesting and lively Crete conference.
This work was supported in part by the US-Israel Binational Science
Foundation, by the German-Israeli Foundation for Scientific Research,
and by the Israel Science Foundation.

\appendix

\section{Bogoliubov transformation}
\label{sec:Bogo}

In the appendix we set our conventions and collect some facts about
Bogoliubov transformations and squeezed states.

\subsection{Transformation of operators}

A Bogoliubov transformation is a linear canonical transformation which mixes
creation and annihilation operators.
It is given by
\begin{equation}
\label{bogTrans}
b_n=W_{nm} a_m+U_{nm} a_m^\dagger\,,\qquad
b_n^\dagger=W_{nm}^* a_m^\dagger+U_{nm}^* a_m\,,
\end{equation}
where $a,a^\dagger$ are the original annihilation and creation operators,
and $b,b^\dagger$ are the new ones.
For this transformation to be canonical, one has to impose
\begin{eqnarray}
[b_n,b_m]=[b^\dagger_n,b^\dagger_m]=0\,\\\qquad
[b_n,b^\dagger_m]=\delta_{n,m}\,.
\end{eqnarray}
These restrictions imply that the matrix $W$ is invertible, the matrix
$S\equiv W^{-1}U$ is symmetric, and that
\begin{equation}
\label{WofS}
(\One-S S^*)=(W^\dagger W)^{-1}\,.
\end{equation}
The normalized vacuum state with respect to the $b$ operators is given by
\begin{equation}
\label{bVac}
\ket{0}_b=\det(\One-S S^*)^{1/4}
    \exp{\left(-\frac{1}{2}a^\dagger S a^\dagger\right)}\ket{0}_a\,,
\end{equation}
where by $a^\dagger S a^\dagger$ we mean the quadratic form
$a^\dagger_n S_{nm} a^\dagger_m$.
The new vacuum is a squeezed state. One can therefore think of the Bogoliubov
transformation as a transformation to a basis where a given squeezed state
plays the role of the vacuum.

Both eq.~(\ref{WofS}),(\ref{bVac}) make sense provided that the eigenvalues of
$S S^*$ (which are necessarily real and positive) are less than 1.
In practice, we will have to deal with the case where this inequality is
saturated. This is the case for the three-vertex, for which all eigenvalues
are $\pm 1$, as well as for projectors, which have eigenvalues $\pm 1$
for $\ka=0$.

Given $S$, eq.~(\ref{WofS}) determines $W$ up to a unitary transformation
that does not alter the vacuum, and under which $W$ transforms as a vector.
Therefore, we can choose $W=W_0$ Hermitian using the Taylor
expansion
\begin{equation}
\label{WHermitian}
W_0=(\One-SS^*)^{-\frac{1}{2}}=\One+\frac{1}{2}SS^*+\frac{3}{8}(SS^*)^2+...\,.
\end{equation}
For a real $S$ we have
\begin{equation}
[W_0,S]=0\,,
\end{equation}
and $W_0$ is real and symmetric.

Bogoliubov transformations can be composed and inverted.
The inverse of transformation~(\ref{bogTrans}) is
\begin{equation}
a=W^\dagger b-U^T b^\dagger\,,
\end{equation}
and given two Bogoliubov transformations
\begin{equation}
b=W_1 a+U_1 a^\dagger\,,\qquad
c=W_2 b+U_2 b^\dagger\,,
\end{equation}
we can write
\begin{equation}
c=(W_2 W_1+U_2 U_1^*)a+(W_2 U_1+U_2 W_1^*) a^\dagger\,.
\end{equation}
It is straightforward to see that this action is associative,
and that the inverse is two sided.
Thus, Bogoliubov transformations form a group with inverse given by
\begin{equation}
\label{InvRule}
(W,U)^{-1}=(W^\dagger,- U^T),
\end{equation}
multiplication rule
\begin{equation}
\label{MultiRule}
(W_2,U_2)\cdot(W_1,U_1)=
(W_2 W_1+U_2 U_1^*,
  W_2 U_1+U_2 W_1^*)\,,
\end{equation}
and identity element $(\One,0)$.

Another useful transformation is to a basis where a given coherent state
plays the role of the vacuum.
\begin{equation}
\ket{0}_b=e^{-\frac{1}{2}\mu^\dagger \mu}
          e^{\mu^T a^\dagger}\ket{0}_a\,,
\end{equation}
where $\mu$ is a given constant vector, and $\ket{0}_b$ is normalized.
It is straightforward to see that
\begin{equation}
\label{coheTrans}
b=a-\mu\,,\qquad
b^\dagger=a^\dagger-\mu^*
\end{equation}
are the canonical operators for which $\ket{0}_b$ is the vacuum.
The inverse transformations is
\begin{align}
a=b+\mu\,,\qquad
a^\dagger=b^\dagger+\mu^*\,,\\
\ket{0}_a=e^{-\frac{1}{2}\mu^\dagger \mu}
          e^{-\mu^T b^\dagger}\ket{0}_b\,.
\end{align}

The transformations (\ref{bogTrans}),(\ref{coheTrans}) can be combined to
form a generalized Bogoliubov transformation whose vacuum is a shifted
squeezed state
\begin{align}
\label{genBogo}
b=&W(a+S a^\dagger-\mu)\equiv W a+U a^\dagger-\sigma\,,\\
\ket{0}_b=&\det(\One-S S^*)^{1/4}
    e^{\frac{1}{2}\Re\left(\mu^\dagger (\One-S S^*)^{-1}S\mu^* \right)-
      \frac{1}{2}\mu^\dagger(\One-S S^*)^{-1}\mu}
    e^{\mu^T a^\dagger-\frac{1}{2}a^\dagger S a^\dagger}
    \ket{0}_a\,.
\end{align}
The composition rule now is
\begin{equation}
\label{genComp}
(W_2,U_2,\sigma_2)\cdot(W_1,U_1,\sigma_1)=
 (W_2 W_1+U_2 U_1^*, W_2 U_1+U_2 W_1^*,W_2 \sigma_1+U_2 \sigma_1^*+\sigma_2)\,,
\end{equation}
the identity is $(\One,0,0)$, and the inverse is
\begin{equation}
\label{genInv}
(W,U,\sigma)^{-1}=(W^\dagger,- U^T,U^T \sigma^*-W^\dagger \sigma)\,.
\end{equation}

\subsection{Transformation of states}
\label{sec:TofStates}

After specifying the transformation of the creation and annihilation
operators and that of the vacuum state, we want to use these results to
calculate the transformation properties of (shifted) squeezed states.
Suppose we are given a shifted squeezed state
\begin{equation}
\ket{V,\mu}=e^{\mu^T a^\dagger-\frac{1}{2}a^\dagger V a^\dagger}\ket{0}_a\,,
\end{equation}
and want to describe it in the basis of $b=W a+ U a^\dagger-\sigma$.
We can think of this state as a ground state (up to the normalization)
after the Bogoliubov transformation
$\mbox{$c=W_V(a+Va^\dagger-\mu)$}$, that is
\begin{equation}
\label{Vof0c}
\ket{V,\mu}=\det(1-V V^*)^{-1/4}
   e^{-\frac{1}{2}\Re\left(\mu^\dagger (\One-V V^*)^{-1}V\mu^* \right)+
  \frac{1}{2}\mu^\dagger (\One-V V^*)^{-1} \mu}\ket{0}_c\,.
\end{equation}
To describe this state in the $b$ basis, we have to compose the
transformation from $a$ to $c$ on the inverse of the transformation from
$a$ to $b$, that is to write
\begin{equation}
\label{0cof0b}
\ket{0}_c=\det(\One-\hat V \hat V^*)^{1/4}
    e^{\frac{1}{2}\Re\left(\hat \mu^\dagger
      \hat (\One-\hat V \hat V^*)^{-1}V\hat \mu^* \right)-
  \frac{1}{2}\hat \mu^\dagger (\One-\hat V \hat V^*)^{-1}
     \hat \mu}e^{\hat \mu^T b^\dagger-\frac{1}{2}b^\dagger \hat V b^\dagger}
    \ket{0}_b\,,
\end{equation}
where, using~(\ref{genComp}),(\ref{genInv}), we find that
\begin{equation}
\label{SofVUW}
\begin{aligned}
\hat V=&(W^\dagger-V U^\dagger)^{-1}(VW^T-U^T)\,,\\
\hat \mu=&(W^\dagger-V U^\dagger)^{-1}
    ((VU^\dagger-W^\dagger)\sigma+(U^T-VW^T)\sigma^*+\mu)\,.
\end{aligned}
\end{equation}
The expression for $\ket{V,\mu}$ is given by combining
eq.~(\ref{Vof0c}),(\ref{0cof0b}),(\ref{SofVUW}).
We see that it does not depend on the (somewhat arbitrary) choice of $W_V$.

\subsection{Transformation of coordinates and momenta}

We can represent the effect of the Bogoliubov transformation on the
canonical coordinates. Given the transformation~(\ref{genBogo}), we define
\begin{equation}
\begin{aligned}
p_a=&\frac{1}{\sqrt{2}}(a+a^\dagger) \qquad
p_b=\frac{1}{\sqrt{2}}(b+b^\dagger) \\
x_a=&\frac{i}{\sqrt{2}}(a-a^\dagger) \qquad
x_b=\frac{i}{\sqrt{2}}(b-b^\dagger)\,.
\end{aligned}
\end{equation}
We could, in principle, allow for $\omega$ dependence here, but that can be
thought of as a composition of yet another \bo transformation.
On the canonical variables the transformation acts as a linear transformation.
\begin{equation}
\label{coorTrans}
\mat{x_b \\ p_b}=\mat{\Re(W-U) & -\Im(W+U) \\
                        \Im(W-U) &\phantom{-} \Re(W+U)}\mat{x_a \\ p_a}+
                \sqrt{2}\mat{\phantom{-}\Im(\sigma) \\ -\Re(\sigma)}\,.
\end{equation}
The non-homogeneous part of it is given by $\sigma$. This is the most general
real linear canonical transformation.

\bibliography{FKM}

\end{document}